\documentclass[10pt,journal,compsoc]{IEEEtran}
\ifCLASSOPTIONcompsoc
  \usepackage[nocompress]{cite}
\else
  \usepackage{cite}
\fi
\ifCLASSINFOpdf
\else
\fi

\usepackage[nocompress]{cite}
\usepackage{subfigure} 
\usepackage{multirow} 
\usepackage{textcomp} 
\usepackage[T1]{fontenc} 
\usepackage{amssymb} 
\usepackage{amsmath,bm} 
\usepackage{graphicx} 
\usepackage{epsfig} 
\usepackage{multirow} 
\usepackage{flushend} 
\usepackage{color}
\usepackage{threeparttable}

\usepackage{amsfonts}
\usepackage{url}
\usepackage{algorithmic}
\usepackage{algorithm}
\usepackage{hyperref}
\newcommand{\x}{\mathbf{x}}

\newcommand{\p}{\mathbf{p}}
\newcommand{\q}{\mathbf{q}}

\usepackage{hyperref}
\usepackage[normalem]{ulem}
\usepackage{color}
\newcounter{ToDo}
\newcounter{gaocomm}
\newcounter{Note}
\definecolor{blue-violet}{rgb}{0.54, 0.17, 0.89}
\definecolor{mygreen}{rgb}{0.0, 0.5, 0.0}
\definecolor{awesome}{rgb}{1.0, 0.13, 0.32}
\definecolor{bostonuniversityred}{rgb}{0.8, 0.0, 0.0}

\newtheorem{definition}{\textbf{Definition}}

\begin{document}
%
\title{A Review for Weighted MinHash Algorithms}
%
%
%
%

\author{Wei Wu,
        Bin Li,
        Ling Chen,
        Junbin Gao 
        and Chengqi Zhang, \emph{Senior Member, IEEE}
\thanks{W. Wu and J. Gao are with the Discipline of Business Analytics, The University of Sydney Business School, Darlington, NSW 2006, Australia, william.third.wu@gmail.com, junbin.gao@sydney.edu.au}
\thanks{B. Li is with the School of Computer Science, Fudan University, Shanghai, China. E-mail: libin@fudan.edu.cn.}
\thanks{L. Chen and C. Zhang are with the Centre for Artificial Intelligence, FEIT, University of Technology Sydney, Ultimo, NSW 2007, Australia. E-mail: \{ling.chen,chengqi.zhang\}@uts.edu.au.}}
\IEEEtitleabstractindextext{%
\begin{abstract}
Data similarity (or distance) computation is a fundamental research topic which underpins many high-level applications based on similarity measures in machine learning and data mining. However, in large-scale real-world scenarios, the exact similarity computation has become daunting due to ``3V'' nature (volume, velocity and variety) of big data. In such cases, the hashing techniques have been verified to efficiently conduct similarity estimation in terms of both theory and practice. Currently, MinHash is a popular technique for efficiently estimating the Jaccard similarity of binary sets and furthermore, weighted MinHash is generalized to estimate the generalized Jaccard similarity of weighted sets. This review focuses on categorizing and discussing the existing works of weighted MinHash algorithms. In this review, we mainly categorize the Weighted MinHash algorithms into quantization-based approaches, ``active index''-based ones and others, and show the evolution and inherent connection of the weighted MinHash algorithms, from the integer weighted MinHash algorithms to real-valued weighted MinHash ones (particularly the Consistent Weighted Sampling scheme). Also, we have developed a python toolbox for the algorithms, and released it in our github. Based on the toolbox, we experimentally conduct a comprehensive comparative study of the standard MinHash algorithm and the weighted MinHash ones.
\end{abstract}

\begin{IEEEkeywords}
Jaccard Similarity; Generalized Jaccard Similarity; MinHash; Weighted MinHash; Consistent Weighted Sampling; LSH; Data Mining; Algorithm; Review
\end{IEEEkeywords}}

\maketitle

\IEEEdisplaynontitleabstractindextext

%
\IEEEpeerreviewmaketitle



\section{Introduction}\label{sec:intro}

%
%
%
%

\IEEEPARstart{D}{ata} are exploding in the era of big data. In 2017, Google search received over 63,000 requests per second on any given day~\cite{google}; Facebook had more than 2 billion daily active users~\cite{facebook}; Twitter sent 500 million tweets per day \cite{twitter} -- Big data have been driving machine learning and data mining research in both academia and industry~\cite{dumbill2013revolution,rajaraman2012mining}. Data similarity (or distance) computation is a fundamental research topic which underpins many high-level applications based on similarity measures in machine learning and data mining, e.g., classification, clustering, regression, retrieval and visualization. However, it has been daunting for large-scale data analytics to exactly compute similarity because of ``3V'' characteristics (volume, velocity and variety). For example, in the case of text mining, it is intractable to enumerate the complete feature set (e.g., over $10^8$ elements in the case of 5-grams in the original data~\cite{rajaraman2012mining}). Therefore, it is imperative to develop efficient yet accurate similarity estimation algorithms in large-scale data analytics.

A powerful solution to tackle the issue is to employ Locality Sensitive Hashing (LSH) techniques~\cite{indyk1998approximate,gionis1999similarity}, which, as a vital building block for large-scale data analytics, are extensively used to efficiently and unbiasedly approximate certain similarity (or distance) measures. LSH aims to map the similar data objects to the same data points represented as the hash codes with a higher probability than the dissimilar ones by adopting a family of randomized hash functions. So far researchers have proposed many LSH methods, e.g., MinHash for the Jaccard similarity~\cite{broder1998min}, SimHash for the cosine similarity~\cite{charikar2002similarity,manku2007detecting}, and LSH with $p$-stable distribution for the $l_p$ distance~\cite{datar2004locality}. Particularly, MinHash has been verified to be effective in document analysis based on the bag-of-words 
model~\cite{shrivastava2014defense} and also, widely used to solve the real-world problems such as social networks \cite{chierichetti2009compressing,wu2018efficient}, chemical compounds \cite{wu2018k} and information management \cite{cherkasova2009applying}. Furthermore, some variations of MinHash have improved its efficiency. For example, $b$-bit MinHash~\cite{li2010b,mitzenmacher2014efficient} dramatically saves storage space by preserving only the lowest $b$ bits of each hash value; while one-permutation MinHash~\cite{li2012one,shrivastava2014densifying} employs only one permutation to improve the computational efficiency.

MinHash and its variations consider all the elements equally, thus selecting one element uniformly from the set. However, in many cases, one wishes that an element can be selected with a probability proportional to its importance (or weight). A typical example is the \emph{tf-idf} adopted in text mining, where each term is assigned with a positive value to indicate its importance in the documents. In this case, the standard MinHash algorithm simply discards the weights. To address this challenge, weighted MinHash algorithms have been explored to estimate the generalized Jaccard similarity~\cite{haveliwala2000scalable}, which generalizes the Jaccard similarity of weighted sets. Furthermore, the generalized Jaccard similarity has been successfully adopted in a wide variety of applications, e.g., malware classification \cite{raff2017malware}, malware detection \cite{drew2016polymorphic}, hierarchical topic extraction \cite{gollapudi2006exploiting}, etc. 

In this review, we give a comprehensive overview of the existing works of weighted MinHash algorithms, which would be mainly classified into \emph{quantization-based}, \emph{``active index''-based} approaches and \emph{others}. The first class aims to transform the weighted sets into the binary sets by quantizing each weighted element into a number of distinct and equal-sized subelements. In this case, the hash value of each subelement is compulsorily computed. By contrast, the second one just considers a couple of special subelements called ``active index'' and compute their hash values, thus improving the efficiency of the weighted MinHash algorithms. The third category consists of various algorithms which cannot be classified into the above categories.

In summary, our contributions are three-fold:
\begin{enumerate}
\item We review the MinHash algorithm and 12 weighted MinHash algorithms, and propose categorizing them into groups.
\item We develop a python toolbox, which consists of the MinHash algorithm and 12 weighted MinHash algorithms, for the review, and release the toolbox in our github\footnote{\url{https://github.com/williamweiwu/williamweiwu.github.io/tree/master/WeightedMinHashToolbox}}.
\item Based on our toolbox, we conduct comprehensive comparative study of the ability of all algorithms in estimating the generalized Jaccard similarity.
\end{enumerate}

The remainder of the review is organized as follows. In Section \ref{sec:oveview}, we first give a general overview for Locality Sensitive Hashing and MinHash, and the categories of weighted MinHash algorithms. Then, we introduce quantization-based Weighted MinHash algorithms, ``active index''-based Weighted MinHash and others in Section \ref{sec:quantization}, Section \ref{sec:sample} and Section \ref{sec:others}, respectively. In Section \ref{sec:exp}, we conduct a comprehensive comparative study of the ability of the standard MinHash algorithm and the weighted MinHash algorithms in estimating the generalized Jaccard similarity. Finally, we conclude the review in Section~\ref{sec:con}.

\section{Overview}
\label{sec:oveview}

\subsection{Locality Sensitivity Hashing}

One fundamental research topic in data mining and machine learning is data similarity (or distance) computation. Based on data similarity, one can further conduct classification, clustering, regression, retrieval and visualization, etc. So far some classical similarity (or distance) measures have been adopted, e.g., $l_p$ distance, cosine similarity, hamming distance and Jaccard similarity, etc. The problem of similarity computation is defined as the nearest neighbor (NN) problem as follows.

\begin{definition}[Nearest Neighbor]
  Given a query object $\q$, the goal is to find an object ${\rm NN}(\q)$, called nearest neighbor, from a set of objects $\mathcal{P}=\{\p_1,\p_2,\cdots,\p_N\}$ so that ${\rm NN}(\q)=\arg\min_{\p\in \mathcal{P}}{\rm dist}(\q,\p)$, where ${\rm dist}(\q,\p)$ is a distance between $\q$ and $\p$.
\end{definition}

Alternatively, there exists a variant of the nearest neighbor problem, i.e., the fixed-radius near neighbor ($R$-NN).
\begin{definition}[$R$-Near Neighbor]
  Given a query object $\q$ and the distance threshold $R$, the goal is to find all objects $R$-${\rm NN}(\q)$ from a set of objects $\mathcal{P}=\{\p_1,\p_2,\cdots,\p_N\}$ so that $R$-${\rm NN}(\q)=\{\p | {\rm dist}(\q, \p)\le R, \p \in \mathcal{P}\}$.
\end{definition}

Finding the nearest neighbor or $R$-near neighbors can be easily solved in the case of low-dimensional data. However, some efficient algorithms for them in low-dimensional spaces will spend high computational cost in the high-dimensional settings due to \emph{curse of dimensionality}. To this end, a large amount of research has concentrated on the approximation problem.

\begin{definition}[$c$-Approximate Near Neighbor]
  Given a query object $\q$, the distance threshold $R>0$, $0<\delta<1$ and $c \ge 1$, the goal is to report some $cR$-near neighbor of $\q$ with the probability of $1-\delta$ if there exists an $R$-near neighbor of $\q$ from a set of objects $\mathcal{P}=\{\p_1,\p_2,\cdots,\p_N\}$.
\end{definition}

One well-known solution scheme for $c$-Approximate Near Neighbor is Locality Sensitivity Hashing (LSH), which aims to map similar data objects to the same data points represented as the hash codes with a higher probability than the dissimilar ones by employing a family of randomized hash functions. Formally, an LSH algorithm is defined as follows:
\begin{definition}[Locality Sensitivity Hashing]
  A family of $\mathcal{H}$ is called $(R, cR, p_1, p_2)$-sensitive if for any two data objects $\p$ and $\q$, and $h$ chosen uniformly from $\mathcal{H}$:
  \begin{itemize}
    \item if ${\rm dist}(\p,\q)\le R$, then ${\rm Pr}(h(\p)=h(\q))\ge p_1$,
    \item if ${\rm dist}(\p,\q)\ge cR$, then ${\rm Pr}(h(\p)=h(\q))\le p_2$,
  \end{itemize}
  where $c>1$ and $p_1>p_2$.
\end{definition}

\begin{table}[t]
  \centering
  \begin{normalsize}
  \caption{Summary of Classical Similarity (Distance) Measures and LSH Algorithms.}
  \fontsize{8pt}{1.2\baselineskip}\selectfont{\begin{tabular}{|l|r|}
  \hline
  Similarity (Distance) Measure     & LSH Algorithm \\
  \hline
  $l_p$ distance, $p\in (0,2]$         &  LSH with $p$-stable distribution \cite{datar2004locality}             \\
  Cosine similarity      & SimHash \cite{charikar2002similarity}       \\
  Jaccard similarity     & MinHash \cite{broder1997resemblance, broder1998min}     \\
  Hamming distance     & [Indyk and Motwani, 1998] \cite{indyk1998approximate}             \\
  $\chi^2$ distance    & $\chi^2$-LSH \cite{gorisse2012locality}     \\
  \hline
  \end{tabular}
  }\label{tab:similarity}
  \end{normalsize}
\end{table}

So far various LSH methods have been proposed based on $l_p$ distance, cosine similarity, hamming distance and Jaccard similarity, which have been successfully applied in statistics \cite{eshghi2008locality}, computer vision \cite{ke2004efficient,zhao2013sim}, multimedia \cite{yu2010combining}, data mining \cite{xiong2015top,yu2017generic}, machine learning \cite{satuluri2012bayesian, neyshabur2015on}, natural language processing \cite{ravichandran2005randomized,ture2011no}, etc. Specifically, MinHash \cite{broder1997resemblance, broder1998min} aims to unbiasedly estimate the Jaccard similarity and has been effectively applied in the bag-of-words model such as duplicate webpage detection \cite{fetterly2003large, henzinger2006finding}, webspam detection \cite{jindal2008opinion,urvoy2008tracking}, text mining \cite{chi2014context, kim2011text}, bioinformatics \cite{wu2018k},  large-scale machine learning systems \cite{li2012one,li2011hashing}, content matching \cite{pandey2009nearest}, social networks \cite{wu2018efficient} etc. LSH with $p$-stable distribution \cite{datar2004locality} is designed for $l_p$ norm $||\x_i-\x_j||_p$, where $p\in (0,2]$. SimHash \cite{charikar2002similarity} is able to give the unbiased estimator for cosine similarity between data objects represented as vectors. For binary vectors, one LSH method \cite{indyk1998approximate} is used to estimate the hamming distance. Besides, Gorisse et al. \cite{gorisse2012locality} propose $\chi^2$-LSH method for the $\chi^2$ distance between two vectors. The above classical similarity (distance) measures and the corresponding LSH algorithms are summarized in Table \ref{tab:similarity}.

\subsection{Jaccard Similarity and MinHash}

The bag-of-words model is widely used in many applications, especially document analysis. In this section, we will introduce the Jaccard similarity, which has been successfully applied in the bag-of-words model \cite{shrivastava2014defense}, and the MinHash algorithm, which is a well-known LSH algorithm and used to unbiasedly estimate the Jaccard similarity. Also, we will introduce their corresponding extended versions.

Given a universal set $\mathcal{U}$ and its subset $\mathcal{S}\subseteq \mathcal{U}$, if for any element $k \in \mathcal{S}$, its weight $S_k=1$ and for any element $k \in \mathcal{U}-\mathcal{S}$, its weight $S_k=0$, then we call $\mathcal{S}$ a binary set; if for any element $k \in \mathcal{S}$, $S_k>0$ and for any element $k \in \mathcal{U}-\mathcal{S}$, its weight $S_k=0$, then we call $\mathcal{S}$ a weighted set. Furthermore, all (sub)sets can be represented as vectors whose lengths are the size of the universal set, for example, $\mathcal{U}=[U_1, U_2, \cdots, U_n]$ and $\mathcal{S}=[S_1, S_2, \cdots, S_n]$. 

\begin{figure}[t]
\centering
\includegraphics[width=\linewidth]{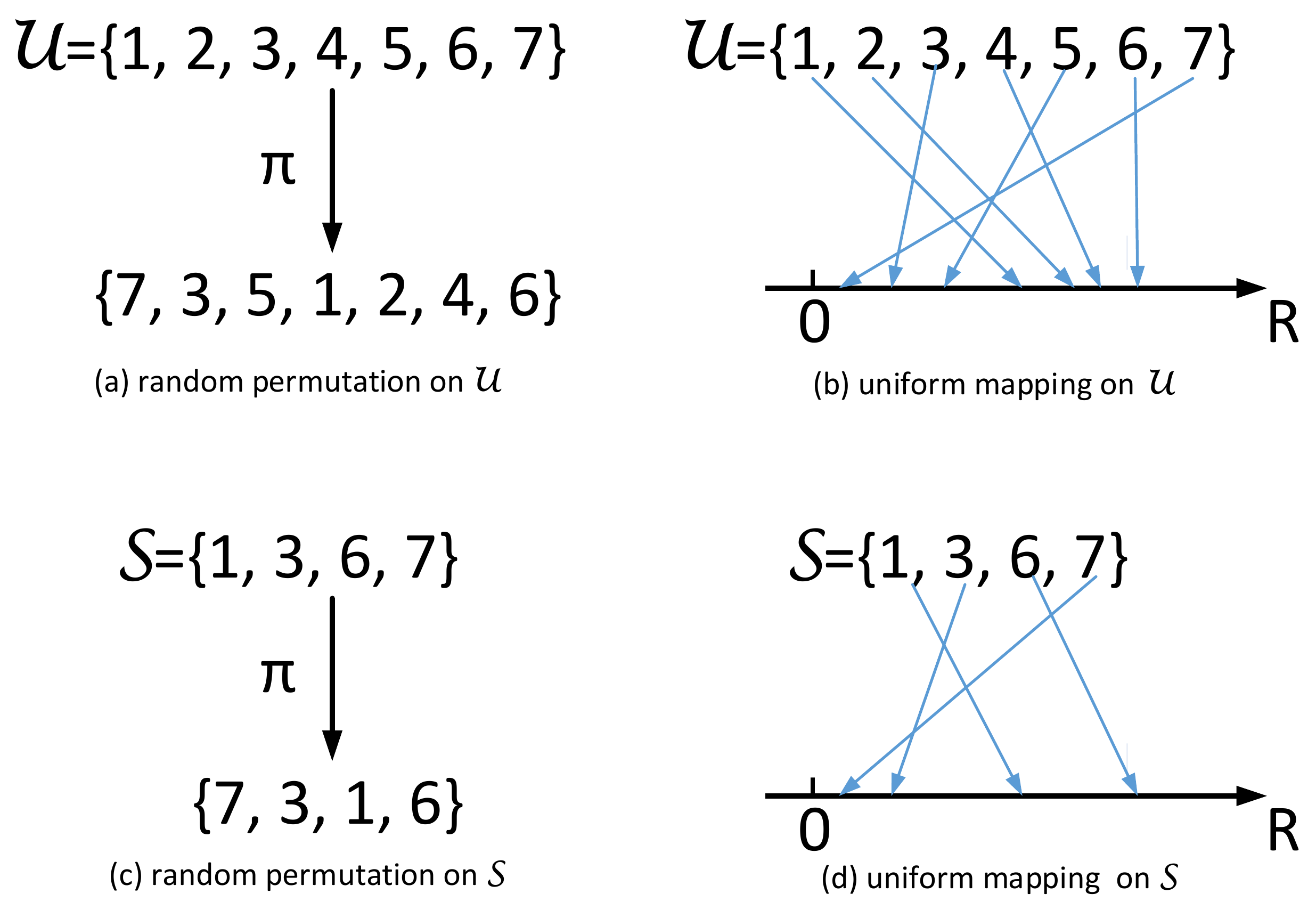}
\caption{A toy example for random permutation and uniform mapping on $\mathcal{U}=\{1,2,3,4,5,6,7\}$ and its subset $\mathcal{S}=\{1,3,6,7\}$. Note that global random permutation or global uniform mapping on the universal set $\mathcal{U}$ and its subset $\mathcal{S}$.}
\label{fig:permutation_mapping}
\end{figure}

A \emph{random} permutation $\pi$ is performed on $\mathcal{U}$, i.e., $\pi: \mathcal{U} \mapsto \mathcal{U}$. Considering the high complexity of the random permutation, alternatively, we can \emph{uniformly} and injectively map each element in $\mathcal{U}$ into the real axis, i.e., $f: \mathcal{U}\mapsto \mathbb{R}$, to approximate the random permutation \cite{gollapudi2006exploiting}. In other words, each element is assigned with a unique hash value $v\in \mathbb{R}$. Figure \ref{fig:permutation_mapping}(a-b) shows the random permutation and uniform mapping on $\mathcal{U}=\{1,2,3,4,5,6,7\}$, respectively. The values on the real axis in Figure \ref{fig:permutation_mapping}(b) imply that the elements are uniformly mapped to the positions in $\mathbb{R}$, which is equivalent to the random permutation in Figure \ref{fig:permutation_mapping}(a).

Furthermore, we can extend uniform mapping on binary sets into weighted sets, i.e., $f: (\mathcal{U},Y) \mapsto \mathbb{R}$, where $Y=[0,S_k]$, which means that each $(U_k, y_k)$, where $U_k\in \mathcal{U}$ and $y_k\in Y$, is distinct. 

Next, we can globally apply random permutation or uniform mapping to a subset $\mathcal{S}$ of $\mathcal{U}$, as shown in Figure \ref{fig:permutation_mapping}(c-d). In other words, the same random permutation or the same uniform mapping is imposed on the universal set and all its subsets. Furthermore, for a subset $\mathcal{S}$, the most important element is the first one in the leftmost, e.g., 7 in Figure \ref{fig:permutation_mapping}(c-d). From the perspective of functions, taking the first element is considered as a hash function, i.e., $h(\mathcal{S})=\min(\pi(\mathcal{S}))$ for random permutation and $h(\mathcal{S})=\min(f(\mathcal{S}))$ for uniform mapping (we adopt $\pi$ to represent the random permutation and uniform mapping for unity below). If we repeat the process $D$ times, we will obtain a fingerprint with $D$ hash values, i.e., $\{\pi_d(\cdot)\}_{d=1}^{D}$.

Now we give two important definitions of the Jaccard similarity and the generalized Jaccard similarity as well as the corresponding solution schemes for the two similarity measures, i.e., the MinHash scheme and the Consistent Weighted Sampling (CWS) scheme.

The Jaccard similarity is statistically a measure of comparing the similarity of two binary sets.
\begin{definition}[Jaccard Similarity]
Given two sets $\mathcal{S}$ and $\mathcal{T}$, the Jaccard similarity is defined as
\begin{equation}
J(\mathcal{S},\mathcal{T})=\frac{|\mathcal{S} \cap \mathcal{T}|}{|\mathcal{S} \cup \mathcal{T}|}.\label{eq:jaccard}
\end{equation}
\end{definition}
The above definition assumes that all elements are equally important. If we extend the definition on weighted sets, we will have the generalized Jaccard similarity as follows.

\begin{definition}[Generalized Jaccard Similarity]
Given two sets $\mathcal{S} = [S_1, \cdots, S_n]$ and $\mathcal{T} = [T_1, \cdots, T_n]$ with all real weights $S_k, T_k \ge 0$ for $k\in\{1,\ldots,n\}$, then the generalized Jaccard similarity is \begin{equation}
generalized J(\mathcal{S},\mathcal{T})=\frac{\sum_{k}\min(S_k,T_k)}{\sum_{k}\max(S_k,T_k)}. \label{eq:generalized_jaccard}
\end{equation}
\end{definition}

It is very time-consuming to compute the (generalized) Jaccard similarity for all paired sets, especially in the case of the large-scale universal set, due to the inefficient operations of union and intersection. To this end, some efficient LSH algorithms have been successively proposed to approximate the similarity measures.

\begin{definition}[MinHash~\cite{broder1998min}]
  Given a universal set $\mathcal{U}$ and a subset $\mathcal{S}\subseteq \mathcal{U}$, MinHash is generated as follows: Assuming a set of $D$ hash functions (or $D$ random permutations), $\{\pi_d\}_{d=1}^D$, where $\pi_d: \mathcal{U} \mapsto \mathcal{U}$, are applied to $\mathcal{U}$, the elements in $\mathcal{S}$ which have the minimum hash value in each hash function (or which are placed in the first position of each permutation), $\{\arg\min(\pi_d(\mathcal{S}))\}_{d=1}^D$, would be the MinHashes of $\mathcal{S}$.
\label{minwise hashing}
\end{definition}

MinHash is used to approximate the Jaccard similarity of two sets. It is proved that the probability of two sets, $\mathcal{S}$ and $\mathcal{T}$, to generate the same MinHash value (hash collision) exactly equals the Jaccard similarity of the two sets $J(\mathcal{S},\mathcal{T})$:
\begin{equation}\label{eq:min}
\Pr[\min(\pi_d(\mathcal{S}))=\min(\pi_d(\mathcal{T}))] = J(\mathcal{S},\mathcal{T}).
\end{equation}

However, it is very expensive for a large universal set to generate the explicit random permutations. Therefore, we generally employ the uniform mapping in practice. To this end, a hash function as follows is adopted to produce the permutated index $\pi_d(i) = (a_d i + b_d) \mod c_d$, where $i$ is the index of an element from the universal set $\mathcal{U}$, $0<a_d,b_d<c_d$ are two random integers and $c_d$ is a big prime number such that $c_d \geq |\mathcal{U}|$ \cite{broder1998min}.

It is easy to see that the MinHash algorithm treats all elements in $\mathcal{U}$ equally because each element can be mapped to the minimum hash value with equal probability. However, if the method is directly used for a weighted set, the weights, which imply the importance level of each element, will have to be simply replaced with 1 or 0. Consequently, the approach will give rise to serious information loss. 

In order to address the aforementioned problem, some weighted MinHash algorithms have been successively proposed, where the Consistent Weighted Sampling (CWS) scheme~\cite{manasse2010consistent} is very remarkable.
\begin{definition}[Consistent Weighted Sampling~\cite{manasse2010consistent}] \label{def:cws}
  Given a weighted set $\mathcal{S} = [S_1,\ldots,S_n]$, where $S_k\ge 0$ for $k\in\{1,\ldots,n\}$, Consistent Weighted Sampling (CWS) generates a sample $(k, y_k): 0 \le y_k \le S_{k}$, which is uniform and consistent.
\end{definition}
\begin{itemize}
\item \textbf{Uniformity:} The subelement $(k, y_k)$ should be uniformly sampled from $\bigcup_{k}(\{k\} \times [0,S_{k}])$, i.e., the probability of selecting the $k$-th element is proportional to $S_{k}$, and $y_k$ is uniformly distributed in $[0, S_{k}]$.
\item \textbf{Consistency:} Given two non-empty weighted sets, $\mathcal{S}$ and $\mathcal{T}$, if $\forall k, T_k \le S_k$, a subelement $(k, y_k)$
is selected from $\mathcal{S}$ and satisfies $y_k \le T_k$, then $(k, y_k)$ will be also selected from $\mathcal{T}$.
\end{itemize}
Similarly, CWS has the following property
\begin{equation}\label{eq:cws}
  \Pr[{\rm CWS}(\mathcal{S})={\rm CWS}(\mathcal{T})] = generalized J(\mathcal{S},\mathcal{T}).
\end{equation}

\begin{table*}[t]
  \centering
  \begin{normalsize}
  \begin{threeparttable}[b]
  \caption{An Overview of Weighted MinHash Algorithms.}
  \fontsize{7.5pt}{1.2\baselineskip}\selectfont{\begin{tabular}{|l|l|l|l|l|}
  \hline
  Category  &  Main Idea &   Algorithms   & Preprocessing & Characteristics\\
  \hline
  \multirow{2}*{Quantization-based}  & \multirow{2}*{Quantized into binary sets}  & [Haveliwala et. al., 2000] \cite{haveliwala2000scalable}     &   Multiply by a large constant & Round off the float part      \\
                                     &  & [Haeupler et.~al.,~2014] \cite{haeupler2014consistent} & Multiply by a large constant       &  Preserve the float part with probability \\\hline
  \multirow{2}*{``Active index''-based} & \multirow{2}*{Sample ``Active index''} & [Gollapudi et. al., 2006](1) \cite{gollapudi2006exploiting}   &   Multiply by a large constant  &   Only sample ``active indices''      \\
  				  & 					& The CWS Scheme \tnote{1}    & - & Extend ``active indices'' to real-value\\\hline
 \multirow{3}*{Others}  &  \multirow{3}*{-}   &                                      [Gollapudi et. al., 2006](2) \cite{gollapudi2006exploiting} & Normalize weights & Preserve elements with probability\\
  &     &
 [Chum et. al., 2008] \cite{chum2008near}   & - & Sample with exponential distribution  \\
  &     & [Shrivastava, 2016] \cite{shrivastava2016simple}  & Require upper bounds of weights & Rejection sampling           \\
  \hline
  \end{tabular}\label{tab:wmh}
  \begin{tablenotes}
     \item[1] See Table \ref{tab:cws}.
   \end{tablenotes}
  }\end{threeparttable} 
  \end{normalsize}
\end{table*}

\begin{table*}[t]
  \centering
  \begin{normalsize}
  \caption{The Algorithms of the CWS Scheme.}
  \fontsize{8pt}{1.2\baselineskip}\selectfont{\begin{tabular}{|l|l|}
  \hline
   Algorithms   & 	Brief Description\\
  \hline
  CWS \cite{manasse2010consistent}   		&  Traverse several ``active indices''\\
  ICWS \cite{ioffe2010improved}   		&  Sample two special ``active indices'' and generate hash code $(k, y_k)$ \\
  0-bit CWS \cite{li20150}  &  Discard $y_k$ produced by ICWS \\
  CCWS  \cite{wu2016canonical}  		&  Uniformly discretize the original weights instead of the logarithm of weighs in ICWS\\
  PCWS \cite{wu2017consistent}   		&  Use one less number of uniform random variables than ICWS\\
  I$^2$CWS \cite{wu2017improved}   	&  Relieve the dependence of two special ``active indices'' in ICWS\\
  \hline
  \end{tabular}
  }\label{tab:cws}
  \end{normalsize}
\end{table*}

\begin{figure*}[t]
\centering
\includegraphics[width=0.8\linewidth]{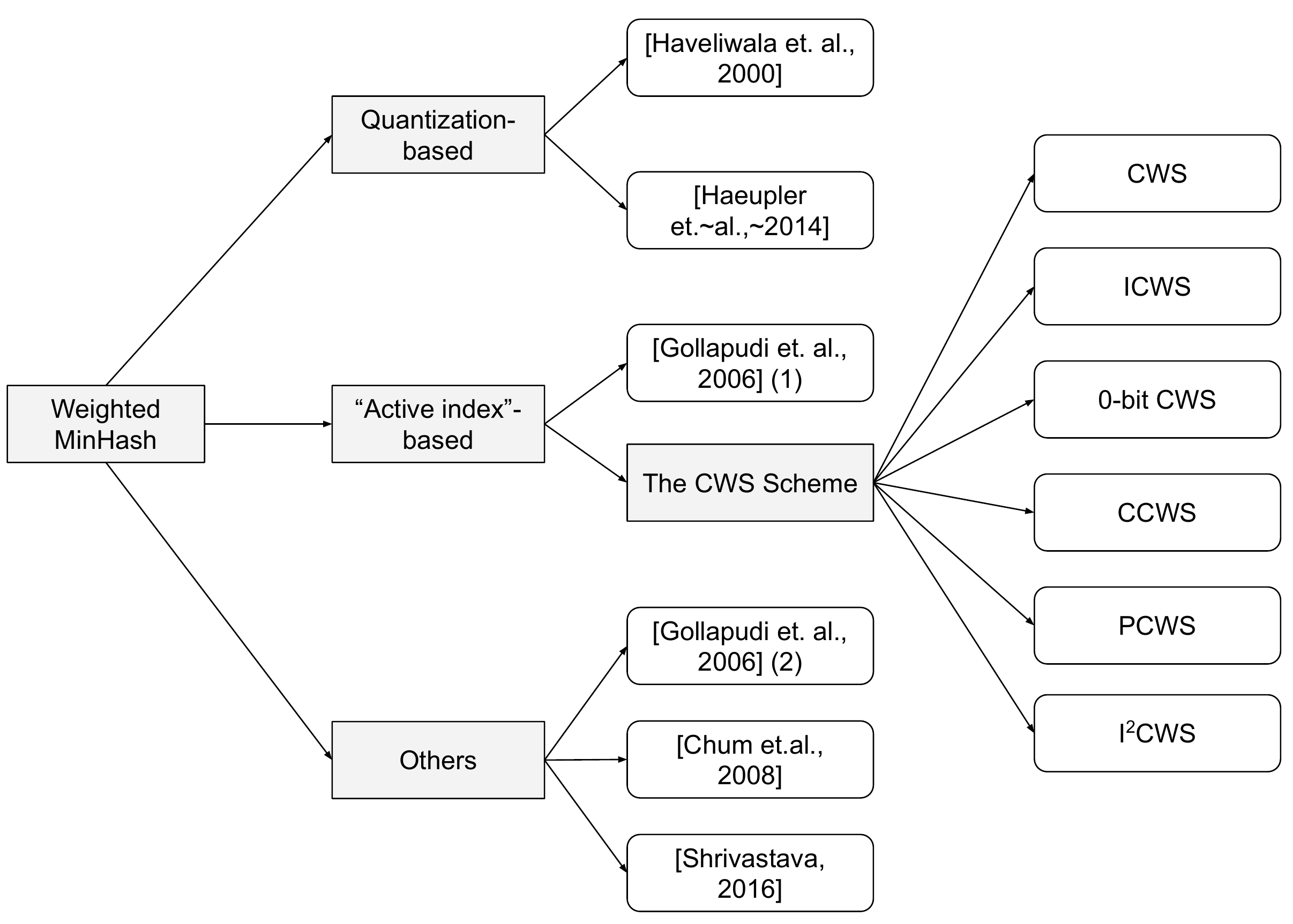}
\caption{An Overview of Weighted MinHash Algorithms}
\label{fig:overview}
\end{figure*}

\subsection{A Categorization of Weighted MinHash Algorithms}

Compared with the MinHash algorithm, the weighted MinHash algorithms preserve the weight information of weighted sets for the sake of performance enhancement. Existing works of weighted MinHash algorithms would be classified into \emph{quantization-based}, \emph{``active index''-based} approaches and \emph{others} based on different modes of processing weights. The algorithms are summarized in Table \ref{tab:wmh}, Table \ref{tab:cws} and Figure \ref{fig:overview}. 

\begin{itemize}
  \item \textbf{Quantization-based}: The methods generate a number of distinct and equal-sized subelements by explicitly quantizing each weighted element. Consequently, the subelements are independently treated in the augmented universal set. Naturally, the computational complexity is proportional to the number of subelements so that the computational cost is unaffordable in the case of numerous subelements. 
  \item \textbf{``Active index''-based}: The methods only compute hash values for special subelements, i.e., ``active indices''. In particular, the CWS scheme has been by far the most successful in terms of theory and practice, and many researchers have successively proposed the follow-up works. Surprisingly, most of the CWS-based algorithms sample only two so-called ``active indices" for each element, thus remarkably decreasing time complexity.
 \item \textbf{Others:} By contrast, the approaches aim to improve efficiency via various techniques as far as possible.
\end{itemize}

\section{Quantization-based Weighted MinHash Algorithms}
\label{sec:quantization}

The key idea of the quantization-based weighted MinHash algorithms is to convert a weighted set into a binary set by explicitly quantizing each weighted element into a number of distinct and equal-sized subelements, e.g., $S_k^{(i)}, i\in\{1,2,\cdots\}$. In other words, each subelement has the length of $\Delta=1$. In the augmented binary set, all subelements are treated \emph{independently}, even if they are from the same weighted elements. Subsequently, the MinHash algorithm will be directly applied to the augmented binary set. Intuitively, the element with a larger weight is quantized into more subelements than the one with a smaller weight, and thus the former can be selected with a higher probability than the latter.

Generally, in order to implement the above conversion, each weight should be multiplied by a large constant $C$ due to the fact that the weights are very small in most cases, e.g., $S_k\in [0,10]$. Consequently, a large $C$ guarantees that the real-valued weights are quantized more precisely and, the accuracy and runtime performance of the algorithms are both closely related to $C$. The accuracy performance gets improved with increasing $C$, but the time complexity is proportional to the size of the augmented universal set. Therefore, it is indispensable to keep a trade-off between accuracy and runtime when this category of algorithms is employed. However, the preprocessing step produces the remaining float part of each weighted element. Naturally, the remaining float part is either preserved or discarded because the MinHash algorithm can only process the existent elements (i.e., $S_k=1$) and the non-existent ones (i.e., $S_k=0$). To this end, two strategies are proposed as follows.

\subsection{[Haveliwala et. al., 2000]}
Haveliwala et. al. (2000) \cite{haveliwala2000scalable} directly rounds off the remaining float part. From the perspective of uniform mapping of MinHash, the algorithm assigns a hash value to each subelement. In this case, the approach can be broken down into two steps: (1) For the $k$-th element in $\mathcal{S}$, assign each subelement $(k, y_{k,i})$ with a hash value and find $(k, y_k)$ that has the minimum hash value; (2) Find $(k, y_k^*)$ that has the minimum hash value in $\{(k, y_k)\}_{k=1}^{n}$.

\subsection{[Haeupler et.~al.,~2014]}
Compared to [Haveliwala et. al., 2000], [Haeupler et.~al.,~2014]~\cite{haeupler2014consistent} preserves the remaining float part with probability being exactly equal to the value of the remaining float part. Specifically, the float part is uniformly and consistently mapped to the interval $[0, 1)$, that is, the random hash value is seeded with the element. If the random value is less than the value of the remaining float part, it will be added as a subelement into the binary set; otherwise, it will be abandoned.

By multiplying a constant $C$, the quantization-based approaches expand the universal set, where each element is permuted or computed for a hash value. 

\section{``Active index"-based Weighted MinHash Algorithms}
\label{sec:sample}

The quantization-based weighted MinHash algorithms need to compute the hash values for all subelements, so it is very inefficient when the universal set is augmented remarkably. In order to address the issue, the researchers proposed the idea of ``active index''~\cite{gollapudi2006exploiting} and many follow-up works, all of which improve the efficiency of the quantization-based weighted MinHash algorithms by just sampling several ``active indices'' and then computing the hash values for them.

\subsection{[Gollapudi et. al., 2006](1)}

Quantization-based weighted MinHash algorithms compute a hash value for each subelement, which is computationally inefficient in practice. In order to accelerate the computation process, Gollapudi and Panigraphy proposed an efficient weighted MinHash algorithm for integer weighted sets, which is able to skip much unnecessary hash value computation by employing the idea of ``active index"~\cite{gollapudi2006exploiting}.

\begin{figure}[t]
\centering
\includegraphics[width=0.8\linewidth]{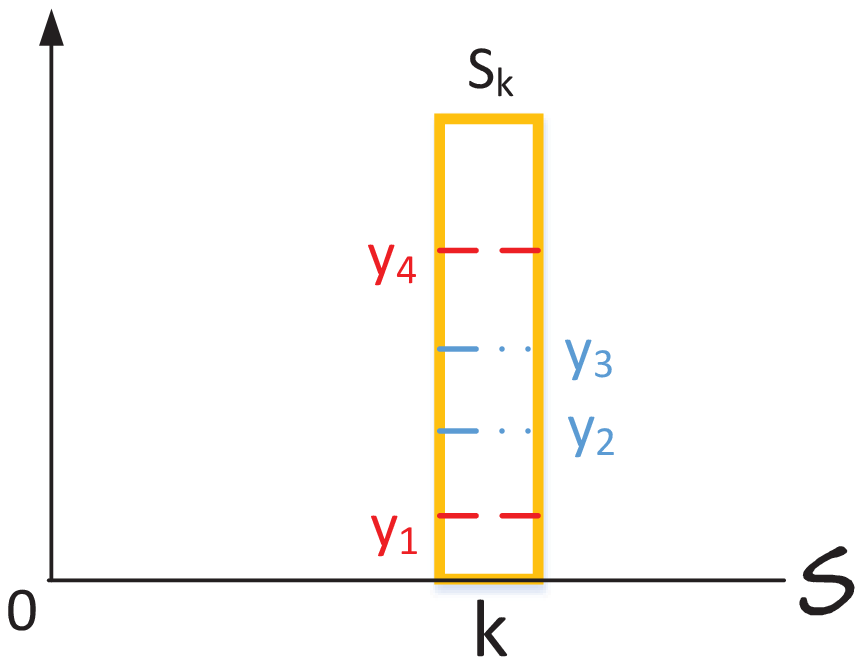}
  \caption{An illustration of integer ``active indices'' on the left side and real-valued ones on the right side. The line squares or rectangle are below the weight $S_k$, while the dash rectangle is above the weight (the same in Figures \ref{fig:consistency} and \ref{fig:icws}). The red squares represent integer ``active indices'' and the red lines denote real-valued ones, whose hash values are monotonically decreasing from bottom to top; the blue squares are integer non-active subelements and the blue areas are real-valued non-active ones. In [Gollapudi et. al., 2006](1), $v_{k,y_{k6}} < v_{k,y_{k4}}< v_{k,y_{k1}}< \{v_{k,y_{k2}}, v_{k,y_{k3}}\}$, $y_{k4}< v_{k,y_{k5}}$ and $v_{k,y_{k6}}< v_{k,y_{k7}}$; In CWS, $v_{k,y_{k15}} < v_{k,y_{k10}} < v_{k,y_{k6}} < v_{k,y_{k4}}< v_{k,y_{k1}}$. Note that $y_k$ and $z_k$ are two special ``active indices": The former denotes the \emph{largest ``active index"} smaller than the weight, while the latter is the \emph{smallest ``active index"} greater than the weight. [Gollapudi et. al., 2006](1) proceeds traversing ``active indices", $y_{k1}, y_{k4}$ and $y_{k6}$, below the weight $S_k$, from bottom to top by sampling a skipping interval from the geometric distribution. By contrast, CWS successively explores the intervals in both directions of the interval $(4,8]$ containing $S_k$ until $y_k$ and $z_k$ are found.}
\label{fig:activeIndex}
\end{figure}

The algorithm explores the weights from bottom to top. In the integer weighted set, the weighted element is first quantized into a sequence of subelements, each of which has the length of $\Delta=1$ (i.e., the weight of each subelement is $\Delta=1$), and then each subelement can be uniformly mapped to the interval $[0, 1]$. Furthermore, we observe an interesting phenomenon that there exists a subsequence which starts with the subelement at the bottom and whose hash values are monotonically decreasing from bottom to top, and the subelements of the subsequence are called as ``active indices''. Considering any interval which is composed of two adjacent ``active indices'' in the subsequence, the hash values of the subelements within the interval are definitely greater than that of the lower endpoint. Consequently, the subsequence can be treated as the Bernoulli trials: ``success'' is the event that the hash value of a certain subelement is less than that of the lower endpoint, and the probability of success is the hash value of the lower endpoint; ``failure'' is the event that the hash value of a certain subelement is greater than that of the lower endpoint. Naturally, the number of trials that must be run in order to achieve success, i.e., the length of the interval, can be modeled as the geometric distribution. Based on the idea, the subelements between two adjacent ``active indices'' can be directly skipped by sampling a skipping interval from the geometric distribution.

Take $S_k$ in the left part of Figure~\ref{fig:activeIndex} as an illustration, where we just consider the part below the weight $S_k$: It computes and compares hash value of each subelement from bottom to top. Assuming that for seven subelements, $(k,y_{k1}), (k,y_{k2}),\cdots,(k,y_{k7})$, we have $v_{k,y_{k6}} < v_{k,y_{k4}}< v_{k,y_{k1}}< \{v_{k,y_{k2}}, v_{k,y_{k3}}\}$, $y_{k4}< v_{k,y_{k5}}$ and $v_{k,y_{k6}} <  v_{k,y_{k7}}$. When the algorithm starts from the subelement at the bottom $y_{k1}$, in the following steps, it will directly reach $y_{k4}$ by skipping subelements $y_{k2}$ and $y_{k3}$, because the blue subelements with larger hash values than $y_{k1}$ can be ignored when one only needs to maintain the minimum hash value $v_{k,y_k}$ for the $k$-th element. Specifically, in the interval $[y_{k1}, y_{k4}]$, the hash value $v_{k,y_{k4}}$ of the right endpoint $y_{k4}$ is uniformly distributed on the interval $[0, v_{k,y_{k1}}]$, and thus the number of subelements within this interval conforms to a geometric distribution with the parameter being $v_{k,y_{k1}}$. Similarly, the algorithm directly goes from $y_{k4}$ to $y_{k6}$ while skipping subelements $y_{k5}$. Since there exist no subelements, whose hash value is smaller than that of $y_{k6}$ between $y_{k6}$ and $S_k$, the algorithm ceases to traverse in the element. Consequently, the method can iteratively sample all the ``active indices'' from bottom to top for the sake of $(k,y_k)$ with the minimum hash value. It is easy to see that $y_k$ is the largest ``active index'' in $\{1,2,\ldots,S_k\}$.

Essentially, the method is not able to change the underlying uniform distribution of the minimum hash value among the subelements generated from the weighted set according to the properties of the Bernoulli trials and the geometric distribution. Therefore, it can be considered as the accelerated version of [Haveliwala et. al., 2000]. We observe from Figure \ref{fig:activeIndex} that the method traverses all ``active indices" within the weight, i.e, $[0,S_k]$, while skipping non-active subelements. Consequently, the method remarkably reduces the computational complexity from $\mathcal{O}(\sum_k S_k)$ to $\mathcal{O}(\sum_k \log S_k)$. However, for real-valued weighted sets, the method still needs an explicit quantization by multiplying a large constant $C$, which brings tremendous overhead in both time and space.


\subsection{The Consistent Weighted Sampling Scheme}

[Gollapudi et. al., 2006](1) is still inefficient for real-valued weighted sets because it is actually a weighted MinHash algorithm for integer weighted sets, and real-valued weights have to be transformed into integer weights by multiplying a large constant $C$, thus increasing the number of ``active indices" which need to be traversed. To this end, by setting the weight of the subelement $\Delta\rightarrow 0$, the Consistent Weighted Sampling scheme \cite{manasse2010consistent} was proposed to address the issue that [Gollapudi et. al., 2006](1) cannot directly handle real-valued weights.

\begin{figure}[t]
\centering
\includegraphics[width=\linewidth]{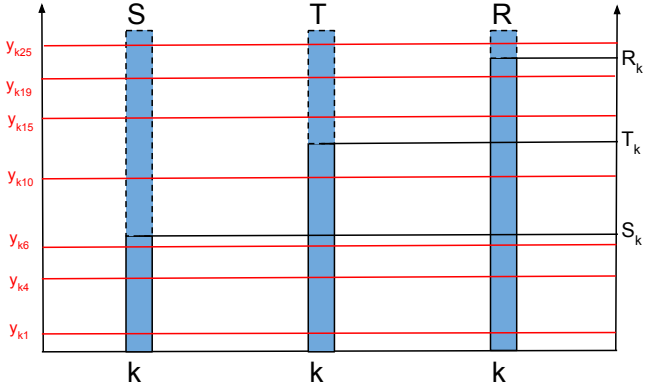}
\caption{A sequence of ``active indices'', e.g., $y_{k1}, y_{k4}, y_{k6}, y_{k10}, y_{k15}, y_{k19}, y_{k25}$, are shared in the $k$-th elements of different weighted sets, e.g., $\mathcal{S}, \mathcal{T}$ and $\mathcal{R}$, and independent of the weights, e.g., $S_k, T_k$ and $R_k$.}
\label{fig:consistency}
\end{figure}

As is known from Definition \ref{def:cws}, the CWS scheme requires two conditions: uniformity and consistency. Uniformity means that each subelement is mapped to the minimum hash value with equal probability. On the other hand, in essence, consistency points out that in a hashing process, the same subelements from different weighted sets are assigned with the same hash value, i.e., global mapping in Figure \ref{fig:permutation_mapping}. If we refer to the decomposition steps of [Haveliwala et. al., 2000], in order to find the ``active index'' of a certain element that has the minimum hash value, a sequence of ``active indices'' are shared in the same elements of different weighted sets and independent of the weights, as shown in Figure \ref{fig:consistency}.

\subsubsection{Consistent Weighted Sampling}
\label{subsec:cws}

The first algorithm of the Consistent Weighted Sampling scheme, the Consistent Weighted Sampling (CWS) algorithm, extends ``active indices" from $[0, S_k]$ in [Gollapudi et. al., 2006](1) to $[0,+\infty)$. As shown in the right part of Figure \ref{fig:activeIndex}, a sequence of ``active indices" are composed of $y_{k1}, y_{k4}, y_{k6}, y_{k10}$ and $y_{k15}$, although $y_{k10}$ and $y_{k15}$ are above the weight $S_k$. It is easy for us to observe two special ``active indices", one of which is the \emph{largest ``active index"} $y_k=y_{k6}\in [0, S_k]$ smaller than the weight and the other of which is the \emph{smallest ``active index"} $z_k=y_{10}\in (S_k,+\infty)$ greater than the weight. In order to find the two ``active indices", instead of starting from 0 in [Gollapudi et. al., 2006](1), CWS explores the ``active indices" in both directions from the weight $S_k$, towards 0 as well as $+\infty$.

Inspired by the fact that the ``active indices'' are uniformly and independently sampled in [Gollapudi et. al., 2006](1), the distribution over ``active indices'' in any interval is independent from the one over ``active indices'' in another disjoint interval. Consequently, CWS partitions $(0, +\infty)$ into intervals of the form $(2^{k-1}, 2^k]$ so that the ``active indices'' can be explored in each interval independently. In other words, one can leap directly to any interval and explore ``active indices'' within it. 

Specifically, in each interval, one starts from the upper endpoint of the interval, and generates a sequence of ``active indices'' from the upper endpoint to the lower one by uniformly sampling. So the ``active indices'' are independent of the specific weight and the sampling operation satisfies the uniformity property of the CWS scheme. In terms of consistency, the ``active indices'' are deduced from the upper endpoints of the intervals, which are shared in the same elements from different weighted sets. Consequently, the two special ``active indices'', $y_k$ and $z_k$, are consistent and paired, which means that the same $y_k$ corresponds to the same $z_k$, and vice versa. In addition, the hash value of $(k, y_k)$ is just related to $z_k$, and thus in the same elements from different weighted set, the same $(k, y_k)$ can be consistently mapped to the same hash value.

CWS starts with the interval containing $S_k$, and if $y_k$ or $z_k$ is not found, it will seek in the next corresponding interval until it is observed. As shown in Figure \ref{fig:activeIndex}, CWS first seeks $y_k$ and $z_k$ in the interval $(4,8]$. $y_k=y_{k6}$ is found, and thus CWS stops traversing towards 0. However, $z_k$ is not in this interval, which makes CWS explore in the next interval $(8,16]$. Consequently, $z_k=y_{10}$ is obtained and CWS stops towards $+\infty$.

CWS runs in expected constant time to find $y_k$ and $z_k$, and thus its average time complexity is $\mathcal{O}(\sum_k \log S_k)$.

\subsubsection{Improved Consistent Weighted Sampling}
\label{subsec:icws}


CWS needs to traverse in both directions for the purpose of the two special ``active indices". Subsequently, Ioffe proposed the Improved Consistent Weighted Sampling (ICWS) algorithm \cite{ioffe2010improved} which can directly sample the two special ``active indices", $y_k$ and $z_k$.

In \cite{ioffe2010improved}, the two special ``active indices" can be obtained via the following equations,
\begin{eqnarray}
  \ln y_k & = & \ln S_k - r_k b_k, \label{uniform2}\\
  \ln z_k & = & r_k+\ln y_k,  \label{eq:rzy}
\end{eqnarray}
where $b_k \sim {\rm Uniform}(0,1)$ and $r_k \sim {\rm Gamma}(2,1)$.

On the other hand, in the algorithmic implementation of ICWS, Eq. (\ref{uniform2}) is replaced with the following one for the sake of consistency
\begin{equation}
  \ln y_k= r_k\left(\left\lfloor \dfrac{\ln S_{k}}{r_k}+ \beta_{k} \right\rfloor - \beta_{k}\right),
\label{uniform1}
\end{equation}
where $\beta_k \sim {\rm Uniform}(0,1)$. It can be proved that $\ln y_k$ in Eq.~(\ref{uniform2}) and $\ln y_k$ in Eq.~(\ref{uniform1}) have the same uniform distribution in $[\ln S_k - r_k, \ln S_k]$. The floor function and the uniform random variable $\beta_k$ in Eq.~(\ref{uniform1}) ensures that a fixed $y_k$ is sampled in an interval of $r_k$. From the perspective of Eq. (\ref{eq:cws}), in this case, ICWS can produce the same $y_k$ (i.e., consistency) if the values of $S_k$ in different weighted sets change subtly, which makes the collision of $(k,y_k)$ possible and the probability unequal to 0. By contrast, in the case of Eq.~(\ref{uniform2}), different $S_k$ definitely generates different $y_k$. In other words, when the weight $S_k$ varies in a certain range, Eq.~(\ref{uniform1}) guarantees that $y_k$ is independent of the weight $S_k$ conditioned on the range and furthermore, Eq.~(\ref{eq:rzy}) makes $y_k$ and $z_k$ paired and consistent. 

\begin{figure}[t]
\centering
\includegraphics[width=\linewidth]{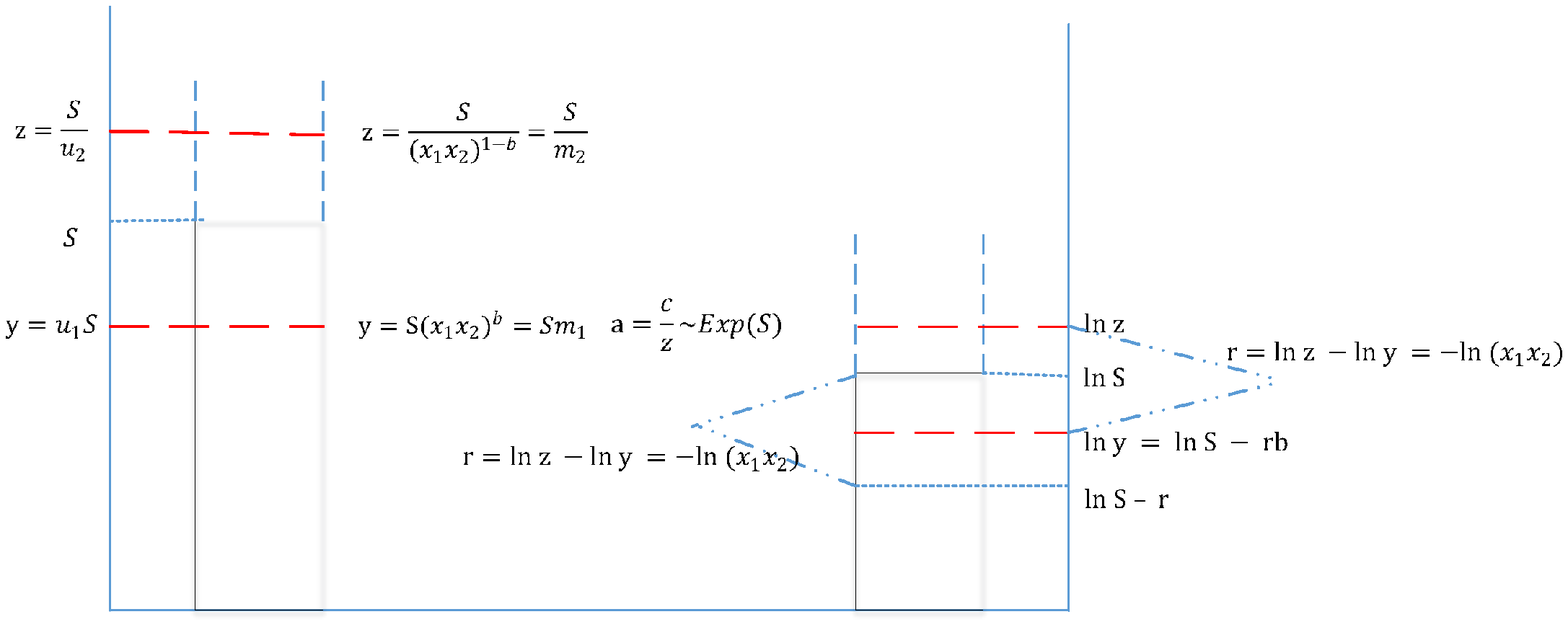}
\caption{A toy example for ICWS. The left bar represents the original weight, and the right one does the logarithm transformation for the weight. The red lines denote two special ``active indices", i.e., $y_k$ and $z_k$.}
\label{fig:icws}
\end{figure}

We know from the above that ICWS implements consistency in a similar way of CWS. In a unified framework, CWS can be quantized through $\frac{\ln S_k}{r_k}$ where $r_k=1$, while ICWS is done through $\frac{\ln S_k}{r_k}$ where $r_k\sim \text{Gamma}(2,1)$. Subsequently, in the two methods, $y_k$ and $z_k$ are actually produced within the intervals which are shared in the same elements from different weighted sets. When the weight $S_k$ fluctuates between $y_k$ and $z_k$, $y_k$ and $z_k$ remain the same, as shown in Figure \ref{fig:icws}. Consequently, it is possible for different weights of each element to generate the consistent $y_k$ and $z_k$.

In order to meet the uniformity of the CWS scheme (i.e., sample $k$ in proportion to $S_k$), ICWS implicitly makes use of a nice property of the exponential distribution: if each hash value $a_{k'}$ of the $k'$-th element is drawn from an exponential distribution parameterized with the corresponding weight, i.e., $a_{k'} \sim {{\rm Exp}(S_{k'})}$, the minimum hash value $a_k$ will be sampled in proportion to $S_k$,
\begin{equation}
  \Pr[a_k=\min\{a_1,\ldots,a_n\}]=\dfrac{S_k}{\sum_{k'} S_{k'}}.
\label{minexp}
\end{equation}

To this end, ICWS implicitly constructs an exponential distribution parameterized with $S_k$, i.e., $a_k\sim {\rm Exp}(S_k)$:
\begin{equation}
a_k = \dfrac{c_k}{z_k}, \label{eq:ayvu}
\end{equation}
where $c_k \sim {\rm Gamma}(2,1)$. In~\cite{ioffe2010improved}, it has been proved that $a_k$ follows the exponential distribution ${\rm Exp}(S_k)$. Based on Eq.~(\ref{eq:ayvu}), the selected $k_*$-th element is returned via $k_* = \arg\min_k a_k$. In addition, it is easy to observe that the hash value of $(k, y_k)$, $a_k$, is unique, and in turn consistency is satisfied.

Essentially, ICWS proceeds the sampling process as follows: It first samples $y_k$ using Eq.~(\ref{eq:yk}), which is derived from Eq.~(\ref{uniform1}). Then the sampled $y_k$, as an independent variable, is fed into Eq.~(\ref{eq:ak}), which is derived from Eqs.~(\ref{eq:rzy}) and~(\ref{eq:ayvu}), and outputs a hash value conforming to the exponential distribution parameterized with the corresponding weight $S_k$.
\begin{eqnarray}
  y_k &= & \exp\left({r_k\left(\left\lfloor \dfrac{\ln S_k}{r_k}+\beta_k \right\rfloor-\beta_k\right)}\right),\label{eq:yk}\\
  a_k & = & \dfrac{c_k}{y_k\exp\left(r_k\right)}.
  \label{eq:ak}
\end{eqnarray}
ICWS calculates the hash value for each element by employing three random variables (i.e., $r_k, \beta_k, c_k$), and the time and space complexity are both $\mathcal{O}(3nD)$ \footnote{In Section \ref{subsec:pcws}, we will demonstrate that ICWS actually adopts five uniform random variables, so it has $\mathcal{O}(5nD)$ in terms of time and space complexity.}.

\subsubsection{0-bit Consistent Weighted Sampling}
\label{subsec:0cws}

The MinHash code produced by ICWS is composed of two components, i.e., $(k,y_k)$. Consequently, there exist two drawbacks: 1) it cannot be integrated into the linear learning systems, which makes it impractical in large-scale industrial applications; 2) $y_k$, as a real value, is theoretically unbounded, which makes it possible for the storage space for $y_k$ to be very large. Consequently, it is not immediately clear how to effectively implement ICWS for building large-scale linear classifiers.

In order to address the above two issues, Li proposed the 0-bit Consistent Weighted Sampling (0-bit CWS) algorithm by simply discarding $y_k$ of the MinHash code $(k,y_k)$ produced by ICWS, that is, 0-bit CWS generates the original MinHash code $(k, y_k)$ by running ICWS, but finally it adopts only $k$ to constitute the fingerprint \cite{li20150}. Although the author empirically demonstrated the effectiveness of the algorithm, a rigorous theoretical proof remains a difficult probability problem.

\subsubsection{Canonical Consistent Weighted Sampling}
\label{subsec:ccws}

\begin{figure}[t]
\centering
\includegraphics[width=\linewidth]{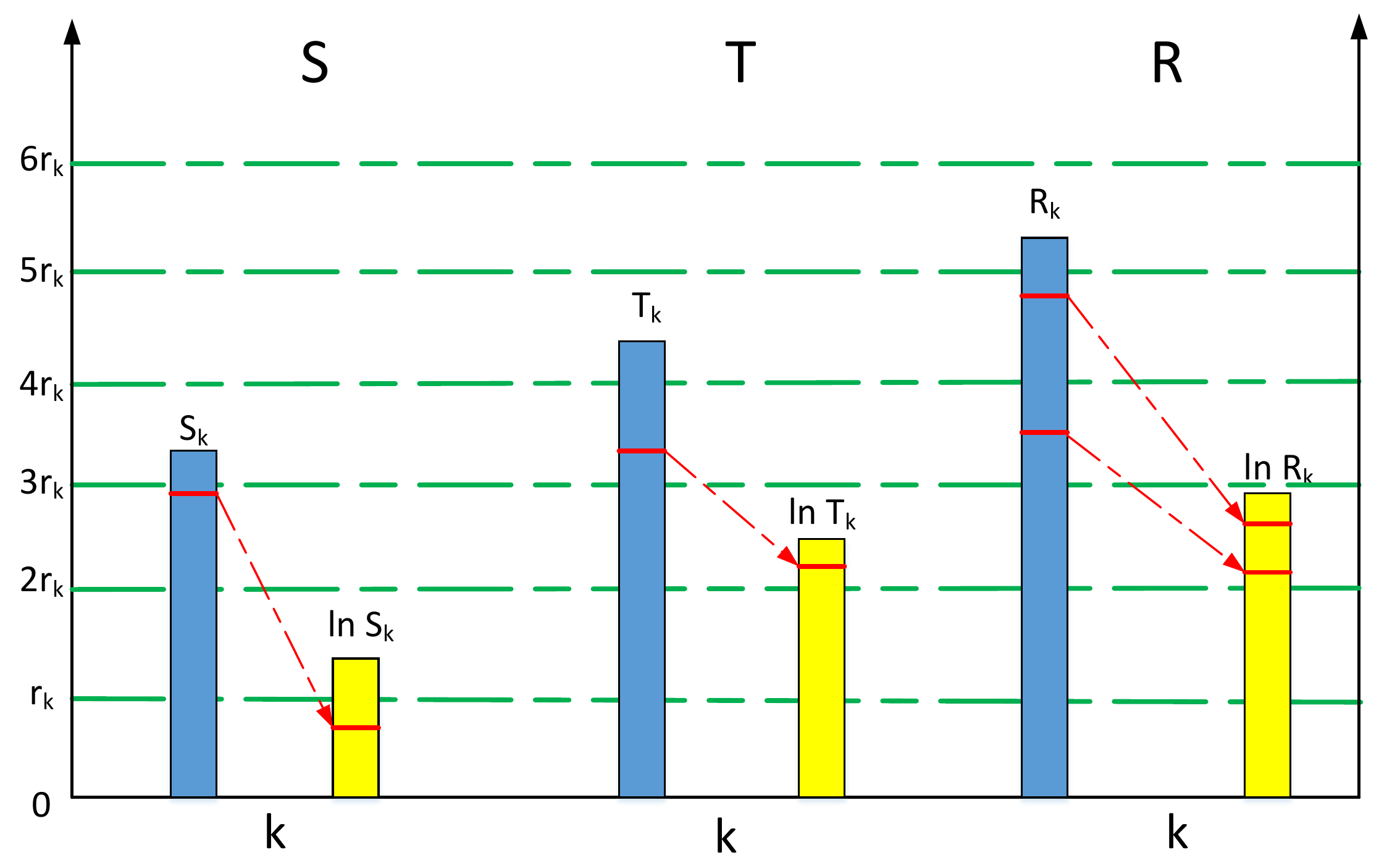}
  \caption{Breach of uniformity. The left bars represent the original weights, while the right ones take the logarithm of the original weights. The red arrows show that the samples are mapped from the original weights into the logarithmic ones. Because of the logarithmic transform, the samples in different subelements of $\mathcal{T}$ and $\mathcal{R}$ are mapped into the same subelement, which increases the collision probability of the ``active indices".}
\label{fig:uniformity_logarithm}
\end{figure}

ICWS directly samples $y_k$ via Eq. (\ref{uniform1}), which essentially conducts an implicit quantization. However, in Eq. (\ref{uniform1}), there exists the logarithm transformation, i.e., $\ln y_k$ and $\ln S_k$, which means that the implicit quantization actually occurs on the logarithm of the original weights and in turn gives rise to non-uniform sampling on the original weights. Consequently, in \cite{wu2016canonical}, Wu et. al. argue that ICWS violates the uniformity property of the MinHash scheme, as shown in Figure \ref{fig:uniformity_logarithm}, and the Canonical Consistent Weighted Sampling (CCWS) algorithm was proposed to avoid the risk of violating the uniformity property of the MinHash scheme.

In CCWS, Eq.~(\ref{uniform2}) is replaced with the following equation
\begin{equation}
y_k=S_k - r_kb_k,
\label{uniform4}
\end{equation}
where $b_k \sim {\rm Uniform(0,1)}$. The above equation effectively avoids scaling the weight sublinearly. Similarly, to implicitly quantize the weights in its algorithmic implementation, Wu et. al. adopt the following equation
\begin{equation}
  y_k= r_{k}\left(\left\lfloor \dfrac{S_{k}}{r_{k}}+ \beta_{k} \right\rfloor - \beta_{k}\right),
\label{uniform3}
\end{equation}
where $\beta_k \sim {\rm Uniform}(0,1)$. Eq.~(\ref{uniform3}) shows that the quantization is directly conducted on the original weight $S_k$. Also, Eq. (\ref{eq:rzy}) is replaced with 
\begin{equation}
    r_k  =  \dfrac{1}{2}(\dfrac{1}{y_k}-\dfrac{1}{z_k}), \label{eq:ccws:ryz}
\end{equation}
where $r_k  \sim {\rm Beta}(2,1)$.

To implement uniformity, CCWS utilizes the property of the exponential distribution, Eq. (\ref{minexp}), and constructs the exponential distribution via Eq. (\ref{eq:ayvu}).

CCWS adopts the same framework of ICWS to satisfy uniformity and consistency: uniformity is implemented via the property of the exponential distribution, and consistency is fulfilled by generating consistent $y_k$ and $z_k$ and uniquely assigning a hash value to $(k, y_k)$.

Although CCWS preserves the uniformity property by uniformly discretizing the original weights, it decreases the probability of collision and thus generally performs worse than ICWS. On the other hand, we would like to note that there exists limitation in the process of quantization: $y_k$ is sampled from $[S_k-r_k, S_k]$ instead of of $[0, S_k]$, which can be appropriately solved by scaling the weight.

\subsubsection{Practical Consistent Weighted Sampling}
\label{subsec:pcws}

ICWS employs three random variables (i.e., $r_k, \beta_k, c_k$) for each element. Due to the fact that random variables are derived from the uniform random ones in programming languages, it essentially generates five uniform random variables for each element, i.e., $r_k=-\ln(u_{k1}u_{k2})\sim \text{Gamma}(2, 1), \beta_k\sim \text{Uniform}(0,1), c_k=-\ln(v_{k1}v_{k2})\sim \text{Gamma}(2, 1)$, where $u_{k1}, u_{k2}, v_{k1}, v_{k2} \sim {\rm Uniform}(0,1)$. Inspired by this, Wu et. al. proposed a more efficient algorithm, the Practical Consistent Weighted Sampling (PCWS) algorithm \cite{wu2017consistent}, by transforming the original form Eq. (\ref{eq:ak}) for sampling $a_k$ in ICWS into an equivalent version and then simplifying it.

Taking into consideration $r_k = -\ln (u_{k1}u_{k2})\sim \text{Gamma}(2, 1)$ and $c_k = -\ln (v_{k1}v_{k2})\sim \text{Gamma}(2, 1)$, where $u_{k1}, u_{k2}, v_{k1}, v_{k2} \sim {\rm Uniform}(0,1)$, Eq. (\ref{eq:ak}) can be rewritten as 
\begin{eqnarray}
a_k & = & \dfrac{-\ln (v_{k1}v_{k2})}{y_k(u_{k1}u_{k2})^{-1}} \nonumber\\
    & = & \dfrac{-\ln (v_{k1}v_{k2})u_{k2}}{y_k u_{k1}^{-1}}. \label{eq:akuv}
\end{eqnarray}

In \cite{wu2017consistent}, Wu et. al. has proved that 
\begin{equation}
    -\ln (v_{k1}v_{k2})u_{k2} \sim {\rm Exp}(1), \label{eq:v1v2u2}
\end{equation}
and 
\begin{equation}
    \mathbb{E}(y_ku_{k1}^{-1})= \mathbb{E}(\hat{S_k}) = S_k, \label{eq:yu1}
\end{equation}
where $\hat{S}_k$ is an unbiased estimator for $S_k$, and $\hat{S}_k = y_ku_{k1}^{-1}$.

Consequently, Eq. (\ref{eq:v1v2u2}) can be simplified as
\begin{equation}
    -\ln x_k \sim {\rm Exp}(1), \label{eq:x}
\end{equation}
where $x_k\sim {\rm Uinform}(0,1)$, and 
\begin{equation}
    a_k = \dfrac{-\ln x_k}{\hat{S_k}} \sim {\rm Exp}(\hat{S}_k). \label{eq:ms}
\end{equation}

Also, Eq.~(\ref{uniform1}) can be rewritten as 
\begin{equation}
	  \ln y_k= -\ln (u_{k1}u_{k2})\left(\left\lfloor \dfrac{\ln S_{k}}{-\ln (u_{k1}u_{k2})}+ \beta_{k} \right\rfloor - \beta_{k}\right) \label{eq:pcws:y}
\end{equation}

The only difference between ICWS and PCWS lies in the specific implementation of uniformity, although they both employ the property of exponential distribution. The former adopts Eq.~(\ref{eq:akuv}), which is equivalent to Eq.~(\ref{eq:ayvu}), while the latter uses Eq.~(\ref{eq:ms}).

We know from Eqs.~(\ref{eq:ms}) and (\ref{eq:pcws:y}) that PCWS just employs four uniform random variables for each element, i.e., $u_{k1}, u_{k2}, \beta_k$ and $x_k$. Obviously, the time and space complexity of PCWS are both $\mathcal{O}(4nD)$, while the time and space complexity of ICWS are essentially $\mathcal{O}(5nD)$. Also, the authors shows that PCWS is more efficient empirically than ICWS.

\subsubsection{Improved Improved Consistent Weighted Sampling}
\label{subsec:i2cws}

ICWS complies with the CWS scheme via the seemingly reasonable proof. However, Wu et. al. pointed out the latent violation of the independence condition of the ``active indices", and in turn proposed the Improved Improved Consistent Weighted Sampling (I$^2$CWS) algorithm which abides by the required conditions of the CWS scheme \cite{wu2017improved}.

Intuitively, in ICWS and its variations (i.e., 0-bit CWS, CCWS and PCWS), Eq.~(\ref{eq:rzy}) or Eq.~(\ref{eq:ccws:ryz}) builds the direct connection between the two special ``active indices", $y_k$ and $z_k$, which means that $z_k$ is derived from $y_k$ instead of independent sampling on weights. By contrast, in CWS, $z_k$ is obtained by independently exploring the intervals above the weight.

In depth, inspired by Section \ref{subsec:pcws}, Eqs. (\ref{uniform2}) and (\ref{eq:rzy}) can be rewritten as 
\begin{alignat}{3}
    y_k & = S_{k}(\exp(-r_{k}))^{b_{k}} &=& S_k(x_{k1}x_{k2})^{b_k}, \label{eq:ysub}\\
    z_k & = \dfrac{S_k}{(\exp(-r_{k}))^{1-b_{k}}} & =& \dfrac{S_k}{(x_{k1}x_{k2})^{1-{b_k}}},\label{eq:zsub}
\end{alignat}
where $r_k = -\ln (x_{k1}x_{k2})\sim \text{Gamma}(2, 1), x_{k1}, x_{k2}, b_k\sim \text{Uniform}(0,1)$.

It can be observed that $y_k$ and $z_k$ are acquired via the shared uniform random variables, $x_{k1}, x_{k2}$ and $b_k$, which is the primary cause of the dependence between $y_k$ and $z_k$, and in turn violates the independence condition of the ``active indices" in the CWS scheme.

In order to address the above issue, I$^2$CWS completely avoids the usage of the shared random variables in Eqs.~(\ref{eq:ysub}) and (\ref{eq:zsub}), and employs the following equations to generate $y_k$ and $z_k$ independently:
\begin{eqnarray}
y_{k} & =& S_{k}(\exp({-r_{k1}}))^{b_{{k}1}}, \label{eq:i2cwsy0} \\
z_k &=& \frac{S_k}{(\exp({-r_{k2}}))^{1-b_{k2}}}, \label{eq:i2cwsz0}
\end{eqnarray}
where $r_{k1}, r_{k2}\sim \text{Gamma}(2,1), b_{k1}, b_{k2}\sim \text{Uniform}(0,1)$. Similarly, Eqs. (\ref{eq:i2cwsy0}) and (\ref{eq:i2cwsz0}) are replaced with, for the sake of consistency,
\begin{eqnarray}
y_{k} &=& \exp \left(r_{k1}\left(\left\lfloor \frac{\ln S_{k}}{r_{k1}}+\beta_{k1} \right\rfloor-\beta_{k1}\right)\right), \label{eq:i2cwsy} \\
z_k &=& \exp \left(r_{k2}\left(\left\lfloor \frac{\ln S_{k}}{r_{k2}}+\beta_{k2} \right\rfloor-\beta_{k2}+1\right)\right), \label{eq:i2cwsz}
\end{eqnarray}
where $\beta_{k1}, \beta_{k2}\sim \text{Uniform}(0,1)$.
Then, Eq.~(\ref{eq:ayvu}) is used to construct the exponential distribution for the sake of uniformity.

Obviously, the two special ``active indices'', $y_k$ and $z_k$, are mutually independent because they are deduced from random variables which are mutually independent.  

Similarly, I$^2$CWS still employs Eq.~(\ref{eq:ayvu}) to obtain the hash values $a_k$ of all $(k, y_k)$ and maintain uniformity. 

Furthermore, note that $a_k$ is just a function of $z_k$, which means that we do not require $y_k$ for each element via Eq.~(\ref{eq:i2cwsy}). Instead, we only need to compute $y_k$ via Eq.~(\ref{eq:i2cwsy}) only once, that is, Eq.~(\ref{eq:i2cwsy}) is not executed until the minimum $a_k$ is acquired and $k_* = \arg\min_k a_k$. In comparison, ICWS computes $y_k$ and $z_k$ for each element, while I$^2$CWS does $z_k$ for each element and $y_{k_*}$ for the selected element.

In I$^2$CWS, uniformity and independence are maintained via the property of exponential distribution and independence deduction, respectively; consistency is implemented by complying with the framework of ICWS.  

Finally, the space complexity is $\mathcal{O}(7nD)$ and the time complexity is $\mathcal{O}(5nD)$.

\subsection{Discussion}

The key idea of this category of methods is ``active indices'' on a weighted element, which are independently sampled as a sequence of subelements whose hash values monotonically decrease. [Gollapudi et al., 2006](1) designed for the integer weighted sets quantizes each weighted element into some subelements, each of which has weight $\Delta = 1$. 

The CWS scheme requires uniformity and consistency. Essentially, the two conditions can be extended to weighted MinHash algorithms including the integer and real-valued versions. Uniformity is implemented by independently and uniformly sampling subelements. Consistency requires that the same subelements are mapped to the same hash value, although they may come from different weighted sets. In other words, a global hash function is applied to all sets so that the same subelements from different sets must correspond to the same value. 

In terms of uniformity, in the case of integer weighted sets, [Gollapudi et. al., 2006](1) models the sampling process as \sout{the} Bernoulli trials, which in turn efficiently samples ``active indices'' according to the geometric distribution. By contrast, in the case of real-valued weighted sets, CWS samples ``active indices'' sequentially from the upper endpoint to the lower one of the interval; while ICWS and its variations (i.e., 0-bit CWS, CCWS, PCWS and I$^2$CWS) sample by employing the property of exponential distribution to meet the condition. Interestingly, if we regard the weight in the real axis as the time axis, the geometric distribution deals with the time between successes in a series of independent trials, while the exponential distribution does with the time between occurrences of successive events as time flows by continuously. Obviously, the exponential distribution is a continuous analogue of the geometric distribution, and in the two distributions, the event (i.e., ``active indices'') occurs independently and with equal probability.

In terms of consistency, ``active indices'' are shared on the same elements from different weighted sets. In order to implement it, [Gollapudi et. al., 2006](1) traverses ``active indices'' from 0, and the geometric distributions are consistent on the same elements from different weighted sets, thus generating a sequence of consistent ``active indices''. On the other hand, CWS seeks ``active indices'' from the interval containing the weight, and proceeds towards 0 and $+\infty$, respectively, while ICWS and its variations (i.e., 0-bit CWS, CCWS, PCWS and I$^2$CWS) directly sample two special ``active indices'' based on the weight. We would like to note that the weights of the same elements from different weighted sets are not fixed. Therefore, if ``active indices'' depends on the specific value of the weight, it will be impossible to generate the same ``active indices'' in the case of different weights. To this end, the CWS scheme computes ``active indices'' based on the endpoints of the intervals which are produced by partitioning $[0,+\infty)$. As a result, the consistent endpoints can derive consistent ``active indices''. Actually, the consistent ``active indices'' must be deduced from the fixed values, for example, [Gollapudi et. al., 2006](1) starts from 0, while the CWS scheme explores from the endpoint of the consistent interval.

\section{Others}
\label{sec:others}

In addition to the above two categories, researchers have subsequently proposed some weighted MinHash algorithms by employing diverse techniques.

\subsection{[Gollapudi et. al., 2006](2)}

The second algorithm in \cite{gollapudi2006exploiting} preserves each weighted element by thresholding normalized real-valued weights with random samples. The original weighted set is lossily reduced to a smaller-scale binary set by sampling a representative subelement for each element. Specifically, each element is mapped to a real value in $[0,1]$: if the value is not greater than the normalized weight, the element will be preserved because a certain subelement corresponding to the element is selected; otherwise, it will be discarded since no subelements corresponding to the element are sampled, which is similar to the process of handling the remaining float part in [Haeupler et.~al.,~2014]. Consequently, the weighted set is transformed into a binary set, which is in turn fed into the MinHash algorithm.

Obviously, the method has to pre-scan the weighted set in order to normalize it and biasedly estimates the generalized Jaccard similarity.

\subsection{[Chum et. al., 2008]}

Chum \emph{et. al.} \cite{chum2008near} proposed a simple weighted MinHash algorithm which combines quantization and sampling. Similar to other quantization-based algorithms, each weighted element can be quantized into a number of distinct and equal-sized subelements, e.g., $S_k^{(i)}, i\in\{1,2,\cdots, S_k\}$ in the case that $S_k$ is a positive integer. Subsequently, each subelement is assumed to be assigned with a hash value, e.g., $f(S_k^{(i)})\sim {\rm Uniform}(0,1)$. Naturally, the MinHash value of the $k$-th element is $h(S_k)=\min\{f(S_k^{(i)})\}$. Furthermore, Chum et. al. derived the distribution of $h(S_k)$:
\begin{equation}\label{eq:hdist}
  {\rm Pr}(h(S_k)\le a) =  1-(1-a)^{S_k}.
\end{equation}
Finally, $h(S_k)$ can be expressed as
\begin{equation}\label{eq:bmvc}
  h(S_k) = \dfrac{-\log x}{S_k}, 
\end{equation}
where $x\sim {\rm Uniform}(0,1)$.

Interestingly, Eq. (\ref{eq:bmvc}) can be directly used for real-valued weighted sets without multiplying a large constant. Also, we observe that the approach essentially samples an element proportionally to its weight because Eq. (\ref{eq:bmvc}) is an exponential distribution with parameter being $S_k$, i.e., $\frac{-\log x}{S_k} \sim {\rm Exp}(S_k)$, as shown in Eq. (\ref{eq:ms}), and also, the sampled subelement is used to supersede the whole element. Apparently, the exponential distribution makes each element sampled proportionally to its weight, so it is unnecessary to deduce Eq.~(\ref{eq:bmvc}) from Eq.~(\ref{eq:hdist}).

The Chum's algorithm, ICWS and its variations all employ the exponential distribution so that each element is selected proportionally to its weight. However, the algorithm only offers a biased estimate to the generalized Jaccard similarity since it cannot satisfy the consistency. Superficially, compared with ICWS and its variations, the algorithm  just returns $k$, that is, no $y_k$, and thus the specific subelement cannot be positioned. Essentially, the sampled subelement does not depend on the interval but the weight, and thus once the weight fluctuates, the sampled subelement will change. This is possibly because each subelement from the same elements might be mapped to different hash values in different weighted sets in the case of no explicit function.


\subsection{[Shrivastava, 2016]}
\label{subsec:rejection}

Shrivastava proposed a simple weighted MinHash algorithm \cite{shrivastava2016simple}, which can unbiasedly estimate the generalized Jaccard similarity by uniformly sampling.

\begin{figure}[t]
\centering
\includegraphics[width=\linewidth]{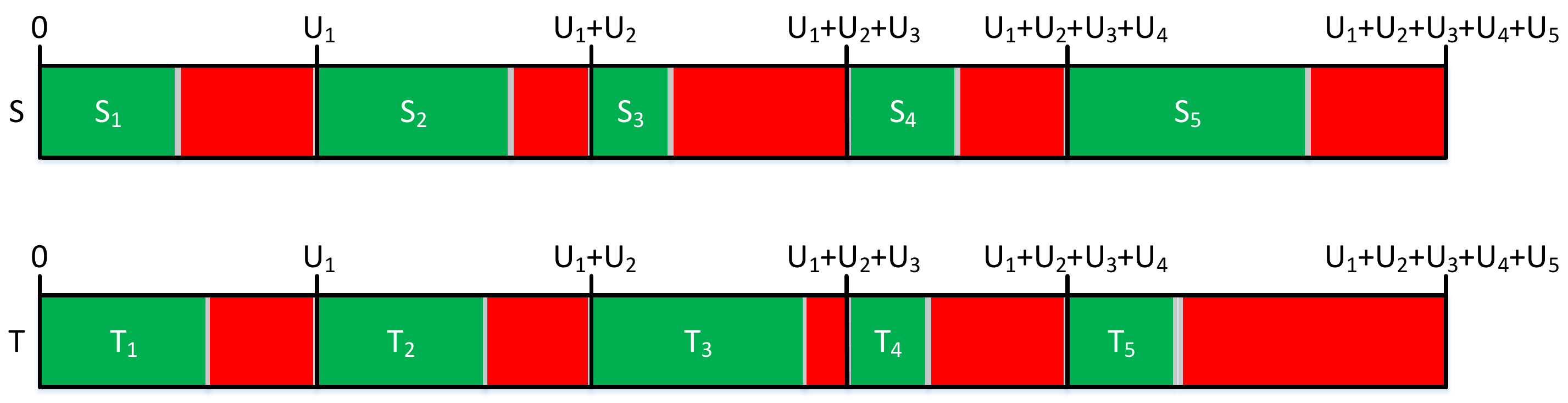}
\caption{A toy example of Red-Green partition on $\mathcal{S}$ and $\mathcal{T}$. $U_i$ represents the maximum weight of each element in the universal set $\mathcal{U}$, and the length of the green area does the weight of each element, i.e., $S_i$ and $T_i$.}
\label{fig:redgreen}
\end{figure}

The algorithm constructs an area in the real axis by concatenating the maximum weight of each element in the data set. As shown in Figure \ref{fig:redgreen}, $U_i$, where $i=\{1,2,3,4,5\}$, represents the maximum weight of the $i$-th element and the length of the $i$-th interval. Naturally, each element in a set can split the corresponding interval into two parts, the green area and the red one. The former denotes the region within the weight, and thus the length is indeed the weight, while the latter does the one outside the weight. The algorithm globally draws an independent and uniform random value from $[0, U_1+U_2+U_3+U_4+U_5]$. If the variable lies in a certain red region, we repeat the above sampling operation until the sample comes from a certain green region. The hash value is the number of steps taken before stopping. Obviously, the global sampling operation guarantees that in one hashing process, all sets share the same sequence of random variables, i.e., consistency, and the same hash values from two sets show that the two samples are both located in the same point within the green area in the real axis, i.e., collision. 

Essentially, the algorithm is based on rejection sampling, where all the green regions represent the true distribution while all the maximum weights of elements constitute a proposal distribution. Consequently, uniform sampling does not stop until the desired sample is acquired, which obviously meets uniformity.

However, we would like to note that it inherits the disadvantage of rejection sampling, which is that if the prefixed upper bound is set very loosely, the acceptance rate will be low and thus the valid samples which are located in the interval $[0, S_k]$ will be difficult to obtain and further give rise to inefficiency. In order to remedy the drawback, we have to acquire the tight upper bound of each element by pre-scanning the whole data set before sampling. Obviously, such a ``pre-scanning'' step increases the workload, especially in large-scale scenarios; it will also prohibit its applications to streaming scenarios where unseen data may exceed the prefixed upper bound. Therefore, the approach is impractical due to the limitation of rejection sampling.

\section{Experiments}
\label{sec:exp}

In this section, we conduct a comparative study on a number of synthetic data sets with power-law distributions to investigate the effectiveness and efficiency of the reviewed methods for the quality of estimators of the generalized Jaccard similarity.

\subsection{Data sets}

The experiments aim to test the efficiency and the bias of estimators, which are calculated by the MinHash algorithm and the weighted MinHash algorithms, for the Jaccard similarity or the generalized Jaccard similarity, by looking into the difference between the estimation result and the real similarity, so toy data are sufficient and we do not need to validate via specific machine learning tasks, e.g., classification or retrieval. In order to simulate the ``bag-of-words" in real text data following power-law distributions, we generate a number of synthetic data sets, each of which contain 1,000 samples and 100,000 features. Similarly, we uniformly produce the dimensions, but the nonzero weights in each vector sample conform to a power-law distribution with the exponent parameter $e$ and the scale parameter $s$. After repeating 1,000 times, we obtain a data set dubbed Syn$e$E$s$S, where ``$e$E" indicates that the exponent parameter of the power-law distribution is $e$ and ``$s$S" indicates that the scale parameter of the power-law distribution is $s$. The data sets are summarized in Table~\ref{tbl:datasets}.

\begin{table*}[t]
\normalsize
\begin{center}
\fontsize{8pt}{1.2\baselineskip}\selectfont
\caption{\normalsize{Summary of the Data Sets}}
\label{tbl:datasets}
\begin{tabular}{|c|r|r|r|r|r|} \hline
Data Set & \# of Docs & \# of Features & Average Density & Average Mean of Weights & Average Std of weights \\\hline
Syn3E0.2S & 1,000 & 100,000 & $0.005$ & 0.2999 & 0.1035\\ \hline
Syn3E0.22S & 1,000 & 100,000 & $0.005$ & 0.3151 & 0.1288\\ \hline
Syn3E0.24S & 1,000 & 100,000 & $0.005$ & 0.3300 & 0.1441\\ \hline
Syn3E0.26S & 1,000 & 100,000 & $0.005$ & 0.3451 & 0.1576\\ \hline
Syn3E0.28S & 1,000 & 100,000 & $0.005$ & 0.3602 & 0.1703\\ \hline
Syn3E0.3S & 1,000 & 100,000 & $0.005$ & 0.3753 & 0.1829\\ \hline
\end{tabular}
\end{center}
\begin{flushleft}
\footnotesize{``\# of Docs'': size of the data set. ``\# of Features'': size of the dictionary (universal set) of the data set. ``Average Density'': ratio of the elements with positive weights to all the elements in the universal set (a small value indicates a sparse data set). ``Average Mean of Weights'': mean of the weights of the documents for each element (a large value indicates that the weights of elements are large). ``Average Std of Weights'': standard deviation of the weights of the documents for each element (a large value indicates that the documents have very different weights for the corresponding element).}
\end{flushleft}
\end{table*}

\subsection{Experimental Preliminaries}
Thirteen state-of-the-art methods, which consist of one standard MinHash algorithm for binary sets and twelve weighted MinHash algorithms for weighted sets, are compared in our experiments:
\begin{enumerate}
\item \textbf{MinHash~\cite{broder1998min}}: The standard Min-Hash scheme is applied by simply treating weighted sets as binary sets;
\item \textbf{[Haveliwala et. al., 2000]~\cite{haveliwala2000scalable}}: This method applies MinHash to the collection of subelements which are generated by explicitly quantizing weighted sets and rounding the remaining float part of the subelements;
\item \textbf{[Haeupler et. al., 2014]~\cite{haeupler2014consistent}}: Compared to [Haveliwala et. al., 2000]~\cite{haveliwala2000scalable}, the method preserves the remaining float part with probability. 
\item \textbf{[Gollapudi et. al., 2006](1)~\cite{gollapudi2006exploiting}}: This method proposes the idea of ``active indices'' and remarkably improves the efficiency of [Haveliwala et. al., 2000]~\cite{haveliwala2000scalable} by skipping a number of ``non-active indices''.
\item \textbf{CWS~\cite{manasse2010consistent}}: This is the first algorithm under the CWS scheme which finds two special ``active indices", i.e., $y_k$ and $z_k$, by traversing intervals.
\item \textbf{ICWS~\cite{ioffe2010improved}}: This method dramatically improves effectiveness and efficiency of the weighted MinHash algorithms by sampling only two special ``active indices'', $y_k$ and $z_k$. 
\item \textbf{0-bit CWS~\cite{li20150}}: This method approximates ICWS by simply discarding one of the two components, $y_k$, in the hash code $(k, y_k)$ of ICWS;
\item \textbf{CCWS~\cite{wu2016canonical}}: Instead of uniformly discretizing the logarithm of the weight as ICWS, this method directly uniformly discretizes the original weight.
\item \textbf{PCWS~\cite{wu2017consistent}}: This approach improves ICWS in both space and time complexities by simplifying the mathematical expressions of ICWS.
\item \textbf{I$^2$CWS~\cite{wu2017improved}}: This method guarantees that the two special ``active indices'', $y_k$ and $z_k$, are sampled independently by avoiding dependence between $y_k$ and $z_k$ in ICWS.
\item \textbf{[Gollapudi et.~al.,~2006](2)~\cite{gollapudi2006exploiting}}: This method transforms weighted sets into binary sets by thresholding real-valued weights with random samples and then applies the standard MinHash scheme;
\item \textbf{[Chum et. al., 2008]~\cite{chum2008near}}: This method essentially samples an element proportionally to its weight via an exponential distribution with the parameter being the weight.
\item \textbf{[Shrivastava, 2016]~\cite{shrivastava2016simple}}: By uniformly sampling the area which is composed of the upper bound of each element in the universal set, this method unbiasedly estimates the generalized Jaccard similarity.
\end{enumerate}

All the compared algorithms are implemented in MATLAB. We first apply all the algorithms to generate the fingerprints of the data. For [Haveliwala et. al., 2000]~\cite{haveliwala2000scalable}, [Gollapudi et. al., 2006](1)~\cite{gollapudi2006exploiting} and [Haeupler et.~al.,~2014]~\cite{haeupler2014consistent}, each weight is scaled up by a factor of 1,000 for the quantization of the subelements. Suppose that each algorithm generates $\x_S$ and $\x_T$, which are the fingerprints with a length of $D$ for the two real-valued weighted sets, $\mathcal{S}$ and $\mathcal{T}$, respectively. The similarity between $\mathcal{S}$ and $\mathcal{T}$ is $$\mathbf{Sim}_{\mathcal{S},\mathcal{T}} = \sum_{d=1}^{D} \dfrac{\mathbf{1}(x_{S,d} = x_{T,d})}{D},$$ where $\mathbf{1}(state) = 1$ if $state$ is true, and $\mathbf{1}(state) = 0$ otherwise. The above equation calculates the ratio of the same MinHash values (i.e., collision) between $\x_S$ and $\x_T$, which is used to approximate the probability that $\mathcal{S}$ and $\mathcal{T}$ generate the same MinHash value, and to estimate the generalized Jaccard similarity. We set $D$, the parameter of the number of hash functions (or random samples), such that $D \in \{10,20,50,100,120,150,200\}$. All the random variables are globally generated at random in one sampling process, that is, the same elements in different weighted sets share the same set of random variables. All the experiments are conducted on a node of a Linux Cluster with $2 \times 2.3$ GHz Intel Xeon CPU (64 bit) and 128GB RAM.

\begin{figure*}[t]
\centering
 \includegraphics[width=\linewidth]{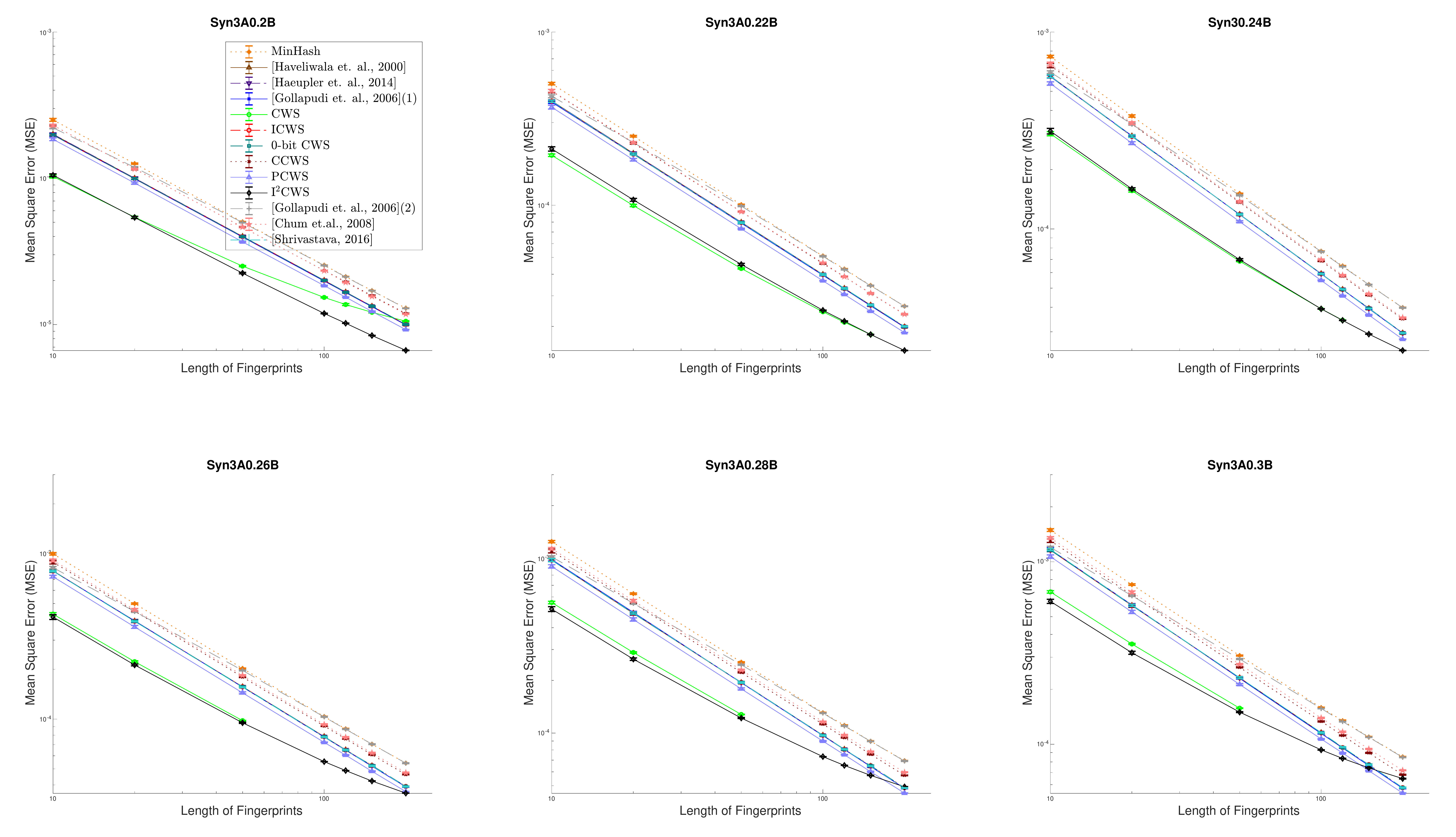}
    \caption{Mean square errors (MSEs) on the synthetic weighted sets with powerlaw distributions. The $x$-axis denotes the length of fingerprints, $D$. Note that in the experiment, each algorithm is given a cutoff of 24 hours. [Shrivastava, 2016] is forced to stop due to running overtime on Syn3A0.2B.}
    \label{fig:accuracy}
\end{figure*}

\begin{figure*}[t]
\centering
\includegraphics[width=\linewidth]{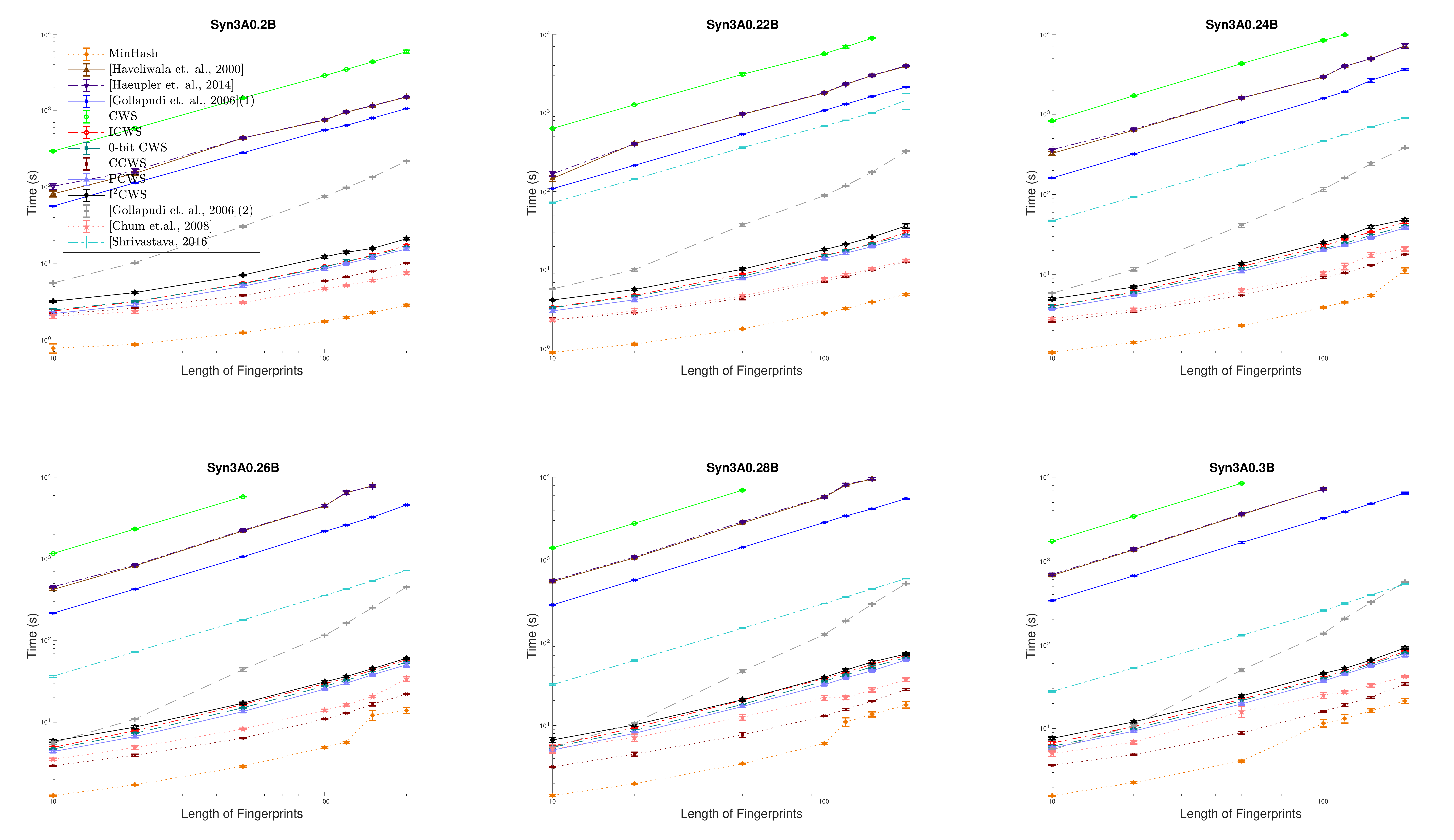}
    \caption{Runtime on the synthetic weighted sets with powerlaw distributions. The $x$-axis denotes the length of fingerprints, $D$. Note that in the experiment, each algorithm is given a cutoff of 24 hours. [Shrivastava, 2016] is forced to stop due to running overtime on Syn3A0.2B.}
    \label{fig:time}
\end{figure*}

\subsection{Experimental Results}
\label{subsec:results}


We empirically present the mean square errors (MSEs) of the estimators of the generalized Jaccard similarity by comparing the estimation results from the compared methods and the real generalized Jaccard similarity calculated using Eq.~(\ref{eq:generalized_jaccard}), and runtime of encoding each data set into fingerprints. We repeat each experiment 10 times and compute the mean and the standard derivation of the results.

Figure \ref{fig:accuracy} and Figure \ref{fig:time} report the mean square errors (MSEs) of the estimators of the generalized Jaccard similarity from 13 compared methods, and their runtime, respectively. Generally, MinHash performs worst but runs faster than most of algorithms because it simply discards the auxiliary weight information of elements. [Haeupler et. al., 2014] performs nearly the same as [Haveliwala et. al., 2000] in terms of accuracy so that they cannot be distinguished. This is largely because the two algorithms both completely preserve the integer parts of elements and the float parts trivially impact the estimators. Also, the two methods maintain the same empirical time level. 

``Active index''-based algorithms not only give more accurate estimators, but also remarkably improve efficiency of the above methods by sampling several and even only two special subelements called ``active indices''. Compared with [Haveliwala et. al., 2000], [Gollapudi et. al., 2006](1) is actually an accelerated version: It transforms a weighted set into a binary set by employing the same preprocessing as [Haveliwala et. al., 2000], i.e, multiplying by a large constant, and then in each element iteratively generates ``active indices'' whose number is less than the value of the augmented weight, thus improving the efficiency of [Haveliwala et. al., 2000] and [Gollapudi et. al., 2006](1), as shown in Figure \ref{fig:time}. Obviously, the time complexity of [Gollapudi et. al., 2006](1) is closely related to the large constant, and it can be further enhanced. It performs the same as [Haveliwala et. al., 2000] and [Haeupler et. al., 2014] in terms of accuracy so that their curves cannot be distinguished in Figure \ref{fig:accuracy}.

By contrast, the Consistent Weighted Sampling (CWS) scheme, as an important category of the real-value weighted MinHash algorithms, can directly handle with real-valued weighted sets. The CWS algorithm, as the first proposed method of the CWS scheme, generally performs best but runs very slowly, possibly because the algorithm iterates many times in order to find the $y_k$ as well as $z_k$ for CWS. Furthermore, the iteration process happens on real values instead of integers, which further increases the number of iterations, because the number of real values to explore is infinite. ICWS performs almost the same performance as 0-bit CWS in terms of accuracy and runtime since one component of ICWS, $y_k$, is trivial to approximate Eq.~(\ref{eq:generalized_jaccard}) for most data sets, which has been experimentally verified in \cite{li20150}. PCWS improves ICWS in terms of accuracy and runtime by adopting one less number of uniform random variables than ICWS. I$^2$CWS obtains similar accuracy with CWS in most cases and meanwhile, defeats other CWS-based algorithms, which demonstrates that I$^2$CWS is able to better approximate Eq.~(\ref{eq:generalized_jaccard}) by fixing the dependence problem of ICWS, but it increases a little time cost. Also, we would like to note that the accuracy performance gain of I$^2$CWS is clear in the case of small $D$, which implies that I$^2$CWS is powerful in the scenarios of limited computational and spatial budget. 

In terms of accuracy, CCWS is inferior to all other CWS-based algorithms because CCWS uniformly discretizes the original weights instead of discretizing the logarithm of the weights in ICWS, 0-bit CWS, PCWS and I$^2$CWS. It is worth noting that the performance of CCWS is getting worse with the increase of the variance of weights of data sets. As shown in~\cite{wu2016canonical}, taking a logarithm on the weights generates an increased probability of collision, and thus CCWS is suitable to data sets with small variance. Surprisingly, CCWS runs more efficiently than ICWS, 0-bit CWS, PCWS and I$^2$CWS because the former directly adopts variables from the {\text Beta} distribution and the {\text Gamma} distribution instead of the {\text Uniform} distribution as the latter do. 



The accuracy gap between [Gollapudi et.~al.,~2006](2) and MinHash narrows down with $D$ increasing, which largely results from the accumulated errors. In terms of runtime, [Gollapudi et.~al.,~2006](2) performs worse because it has to scan the set twice. 
[Shrivastava, 2016](2), which has been proved to exactly estimate Eq.~(\ref{eq:generalized_jaccard}), shows good performance in terms of accuracy and runtime except for Syn3A0.2B. The high efficiency is due to the fact that [Shrivastava, 2016](2) only makes use of a small number of samples, while it fails on Syn3A0.2B, possibly due to the small weights. In this case, small weights make the value of $s_x=\frac{\sum_{k=1}^{n}S_k}{\sum_{k=1}^{n}\max(S_k)}$\footnote{The formula is in Section 5 of \cite{shrivastava2016simple}.} small, where $\max(S_k)$ represents the maximum weight (i.e., the upper bound) of each element in the whole data set, thus decelerating the algorithm. However, we would like to note that it is infeasible to obtain the upper bound of the weight of each element in practice, and thus the method is limited in applying to the real-world problems. [Chum et. al., 2008] performs worse than most of the weighted MinHash algorithms since it just biasedly estimates Eq.~(\ref{eq:generalized_jaccard}), but it runs very fast by simply computation for each element, i.e, Eq.~(\ref{eq:bmvc}).


\section{Conclusion and Future Work}
\label{sec:con}

In this paper, we have reviewed the existing works of weighted MinHash algorithms. The weighted MinHash algorithms can be categorized into quantization-based and ``active index''-based approaches and others.

Currently, most of algorithms come from ``active index''-based methods because they generally improve efficiency by sampling special subelements called ``active indices''. Furthermore, the CWS scheme is remarkable in both theory and practice, and considered as the state-of-the-art methods. Particularly, ICWS and its derivation keep good balance between accuracy and runtime and furthermore, I$^2$CWS and PCWS perform best in terms of accuracy and runtime, respectively \footnote{As mentioned in Section \ref{subsec:results}, CCWS runs efficiently because some variables are directly sampled from the complicated distributions, and thus it is not considered here.} Therefore, they can be applied in large-scale data analytics.

Furthermore, the above methods are designed for static data. However, streaming data, where the elements of a data instance are continuously received in an arbitrary order from the data stream, have been becoming increasingly popular in a wide range of applications such as similarity search in a streaming database and classification over data streams, etc. In this case, some important research issues need to be addressed. First, how to efficiently encode the expandable feature space? The feature space is rapidly expanding due to the fact that the elements continuously arrive. [Gollapudi et. al., 2006](2) and [Shrivastava, 2016] need to construct the complete feature space by pre-scanning, so they are infeasible in streaming data. ICWS and its variations are good solutions, but they require a set of random variables for each element. [Chum et. al., 2008] is efficiently applied to the problem of user activity streams in \cite{yang2016poisketch}. Although it is efficient because only one random variable is required for each element, it cannot unbiasedly estimate the Jaccard similarity, and give the error bound in terms of theory, either. An interesting challenge is whether we are able to further improve the CWS scheme for the purpose of memory-efficiency.

In addition, concept drift should be considered in streaming data because the underlying distribution of streaming data changes over time. The above methods assume that the historical data and the latest data are equally important. However, in the real scenarios, we wish that the algorithms pay more attention to the latter than the former because the latter are commonly more important than the former. So far HistoSketch \cite{yang2017histosketch} has been proposed to solve the problem by employing the strategy of gradual forgetting. Therefore, it is worth exploring a combination of the weighted MinHash algorithms and various strategies of conquering concept drift.

\ifCLASSOPTIONcaptionsoff
  \newpage
\fi



%

%

\bibliographystyle{IEEEtran}
\bibliography{main}
\end{document}